\documentclass[useAMS,usenatbib,asm]{mn2e}
\usepackage{graphicx,fleqn,times,color}
\newcommand{\tal}{\it et al. \rm}

\title{Rings and spirals in barred galaxies. II. Ring and spiral morphology.}
\author[Athanassoula \tal]{E. Athanassoula$^1$, M. Romero-G\'omez$^{1,2}$, A. Bosma$^1$, J.J. Masdemont$^3$ \\ 
$^1$Laboratoire d'Astrophysique de Marseille (LAM), UMR6110, 
CNRS/Universit\'e de Provence,\\
Technop\^ole de Marseille Etoile, 38 rue Fr\'ed\'eric Joliot Curie,
13388 Marseille C\'edex 13, France\\
$^2$Dep. Astronomia i Meteorologia, Universitat de Barcelona, Marti i Franques
1, 08028 Barcelona, Spain\\
$^3$I.E.E.C \& Dep. Mat. Aplicada I, Universitat Polit\`ecnica de
Catalunya, Diagonal 647, 08028 Barcelona, Spain\\
}
\date{Received }

\begin{document}

\maketitle

\begin{abstract}
In this series of papers, we propose a
theory to explain the formation and properties of rings and spirals in
barred galaxies. The building blocks of these structures are orbits
guided by the manifolds emanating from the unstable Lagrangian points
located near the ends of the bar. In this paper we focus on a
comparison of the morphology of observed and of theoretical spirals
and rings and we also give
some predictions for further comparisons. Our theory can account for 
spirals as well as both inner and outer rings. The model outer
rings have the observed $R_1$, $R_1'$, $R_2$, $R_2'$ and  $R_1R_2$
morphologies, including the dimples near the direction of the bar
major axis. We explain why the vast majority of spirals in barred
galaxies are two armed and trailing, and discuss what it would take for
higher multiplicity arms to form. We show that the shapes of observed
and theoretical spirals agree and we predict that stronger
non-axisymmetric forcings at and somewhat beyond corotation will
drive more open spirals. We compare the ratio of ring diameters in
theory and in observations and predict that more elliptical rings will
correspond to stronger forcings. We find that the model
potential may influence strongly the numerical values of these ratios.
\end{abstract}


\section{Introduction}
\label{sec:intro}

The discs of barred galaxies very often have spectacular
sub-structures such as grand design spirals, or inner and outer
rings. In fact, these sub-structures are observed 
much more often in barred than in non-barred disc galaxies. 
Their formation is intimately linked to the evolution of disc
galaxies and also gives important information on the corresponding
underlying potential and the bar pattern speed. Therefore many
efforts, both theoretical and numerical, have been made to explain
them (see e.g. \citealt{Lin67}, \citealt{Toomre77}, \citealt{Toomre81},
\citealt{Atha84}, \citealt{Atha.Bosma.85} and \citealt{Buta.Combes96}
for reviews). None of these theories, however, could 
explain all observed properties and, particularly, none could explain both
spirals and rings. 

For this reason, we propose in this and in previous papers a new
theory to explain all these features in a single theoretical
framework. It relies on the dynamics of the unstable Lagrangian points
located near the ends of the bar and of the associated
manifolds. These guide the motion of the chaotic orbits 
that constitute the building blocks of the spirals and
rings. Our theory and its applications are described in five
papers. The first two (\citealt{RomeroGMAG06}, hereafter Paper I;
\citealt{RomeroGAMG07}, hereafter Paper II) contain most of the
theoretical groundwork. They describe in detail the
dynamics of the region around the two unstable Lagrangian points and
of the corresponding manifolds. A theory, however, can be dynamically
correct but still irrelevant to a particular application. Thus, the aim
of the remaining three papers is to check whether our theory is applicable to
observed spirals and rings. In Paper III \citep{AthaRGM09}, we
considered the orbits guided by the manifolds as potential building
blocks for spiral and rings and derived their relevant properties. In
this paper, the fourth of this series, we   
compare the morphology of observed and of theoretical spirals and
rings to test whether there is agreement.
We also make a number of predictions based on our theory,
which can be tested with future observations. In Paper V (Athanassoula
et al 2009b, to be submitted), the last of this series, we will present
further comparisons with observations. These last three papers can be
considered as one entity and therefore have the same main title (``Rings and
spirals in barred galaxies''), but we will always
consider them as part of the whole series and refer to them as Papers
III, IV and V, respectively. 

The comparisons between our theory and the observations have, by
necessity, to be qualitative. Indeed, in all our 
calculations we have used three specific families of analytic
potentials. These can not cover all possible forms of observed barred
galaxy potentials, nor are we sure that they share all their relevant
properties. Furthermore, we do not take into account  self-gravity,
i.e. we neglect the potential of the structure itself. If 
a given quantity depends considerably on the properties of
the potential, we may not find good quantitative agreement
between our theoretical results and observations, even though our
theory is correct and applicable. For example, 
if the ring extent depends strongly on the potential used, then one
can expect discrepancies between the observed and the theoretical
extent values, if the adopted potential does not match observations in
the relevant regions. 

Section \ref{sec:theory} contains a brief reminder of the main
aspects of our theory. This is necessarily simplified as well as 
condensed, so we refer to an interested reader to Papers I, II
and III and to \citealt{RomeroGMGA09} 
for more information, a more thorough description of the
theory and references to other related work. Nevertheless, the main
theoretical concepts and definitions are given so as to make this paper as
stand-alone as possible, while avoiding repetition. In particular we
introduce here the 
Lagrangian points, the invariant manifolds and the properties of the
associated orbits. Section~\ref{sec:morpho} compares the 
morphological types of our theoretically predicted density enhancements with
the observed ones. Section~\ref{sec:spiralprop} focuses on the
properties of the theoretical and the observed spirals, such as
their sense of winding, their number of arms and their pitch
angle. Section~\ref{sec:ringprop} is devoted to ring properties, such as the
frequency of the various types of rings, the axial ratios of inner and
of outer rings and the ratio of outer to inner ring major
axes. Further comparisons with observations, including pattern speed
prediction, the bar shape, kinematics, radial drift
and abundance gradients will be discussed in Paper V, the last of this
series. There we also give a global discussion of all the results in
this series. For this reason we summarise only briefly the results of
this paper in Sect.~\ref{sec:summary}.
 
\section{Theoretical reminders}
\label{sec:theory}

Our theory is largely based on the dynamics of the Lagrangian points
$L_1$ and $L_2$ of a two-dimensional barred galaxy system. These are
located along the direction of the bar 
major axis and are, in the standard case, unstable saddle points
\citep{Binney.Tremaine08}. Each of them is surrounded by a family of
periodic orbits, called Lyapunov orbits \citep{Lyapunov49}. Since
these are unstable they can not trap around them quasi-periodic orbits
of the same energy\footnote{We will all through this paper loosely
  call `energy' the numerical value of the Hamiltonian in a frame of
  reference co-rotating with the bar, i.e. a
  frame in which the bar is at rest, and denote it by $E_J$.}, so that
any orbit in their 
immediate vicinity (in phase space) will have to escape the
neighbourhood of the corresponding Lagrangian point. Not all departure
directions are, however, possible. The direction in which the orbit
escapes is set by what we call the invariant manifolds. These can be
thought of as tubes that guide the motion of particles of the same
energy as the manifolds \citep{GomezKLMMR, KoonLMR}. Each manifold has
four branches emanating 
from the corresponding Lyapunov orbit, two of them inside corotation
(inner branches) and two of them outside it (outer branches). Along
two of these branches (one inner and one 
outer) the mean motion is towards the region of the Lagrangian point
(stable manifolds), 
while along the other two it is away from it (unstable manifolds). We
need to stress that the terms `stable' and `unstable' do not
mean that the orbits that follow them are stable and unstable,
respectively. In fact all the orbits that follow the manifold are
chaotic, but they are in a loose way `confined' by the manifolds, so
that they stay together, at least for a few bar rotations, in what
could be called a bundle. We propose that these manifolds and orbits
are the building blocks of the spirals and rings.  
 
These manifolds do not exist for all values of the energy, but only for
energies for which the corresponding Lyapunov periodic orbit is
unstable. This means energies within a range starting from the energy
of the $L_1$ or $L_2$ ($E_{J,L_1}$) and extending over a region whose extent
depends on the model \citep{SkokosPA02}.

\begin{figure*}
\centering
\includegraphics[scale=0.5,angle=-90.]{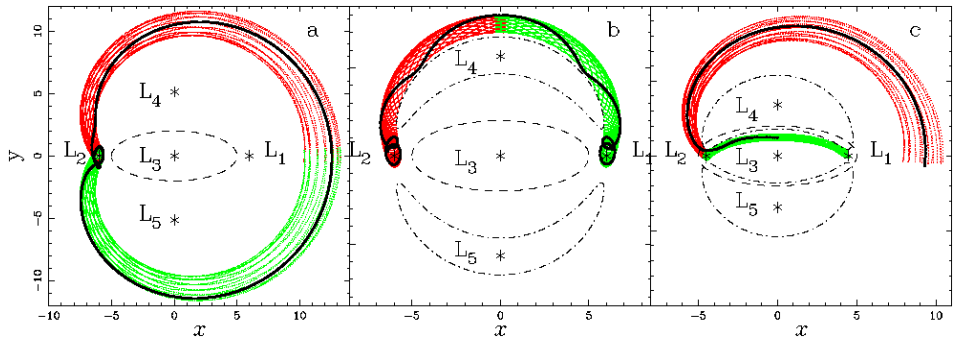}
\caption{Examples of homoclinic (panel a), heteroclinic (panel b) and
  escaping 
(panel c) orbits (black thick lines) in the configuration space. In
red lines, we plot the unstable invariant manifolds associated to these
orbits, while in green we plot the corresponding stable invariant
manifolds. In dashed lines, we give the outline of the bar and, in 
panels (b) and (c), we plot the zero velocity curves in
dot-dashed lines.  
}
\label{fig:homoheterosp}
\end{figure*}

In Papers II and III we classified the manifolds and the corresponding
orbits in three categories, the homoclinic, the heteroclinic and the
escaping ones. Homoclinic manifolds and orbits start from the vicinity
of either the $L_1$, or the $L_2$ and end near the same Lagrangian
point. An example is given in the left panel of Fig.~\ref{fig:homoheterosp}. 
Heteroclinic manifolds and orbits start from either near the
$L_1$, or the $L_2$ and end near the Lagrangian point located at the
other end of the bar, i.e. the $L_2$ or $L_1$, respectively. An example is given
in the middle panel of Fig.~\ref{fig:homoheterosp}. Finally, escaping
orbits start off near one of these Lagrangian points but do not end near
either this point or the symmetric one at the other side of the bar. Their
name comes from the fact that they escape the vicinity of the bar 
and reach parts of the galactic disc further out, but this does {\it not} imply
that they escape the system and reach infinity. 

The case described so far has five Lagrangian points in total, two
unstable and three stable, and is referred to as the standard
case. The formation of the manifolds, and therefore of the rings and
spirals, depends on the existence of such unstable saddle Lagrangian points as
described above. We have, therefore, applied our theory to barred
galaxies. It can. however, be applied also to non-barred galaxies with
other non-axisymmetric perturbations, provided these have unstable
saddle Lagrangian points as described above and in Papers I, II and
III. This can include spiral perturbations and will be discussed
further in Paper V. 

Another alternative, which is interesting both dynamically and
morphologically, occurs in barred galaxies when the $L_1$ and $L_2$ are
stable. This can happen if a sufficiently important amount
of mass is sufficiently concentrated around the Lagrangian points, as
described in Paper III. In such cases, the $L_1$ and $L_2$ become
stable, while four other Lagrangian points appear along the direction
of the bar major axis, one on either side of the $L_1$ and $L_2$.
These new Lagrangian points are unstable and thus
work like the $L_1$ and $L_2$ in the standard case. The intricacies of
the manifolds in such cases have been studied in Paper III, and we
will only mention here that the gross global morphology of the system
stays roughly unchanged for a wide range of parameter values. Other
things. however, such as 
the axial ratio of the rings can change, as will be discussed in
Sect.~\ref{sec:ringprop}.    

In Papers II and III, we calculated manifolds and orbits in three
different rigid barred galaxy potentials. Model A is taken from
\cite{Atha92a} and has a Ferrers bar \citep{Ferrers77}
of semi-major axis $a$, axial ratio $a/b$, quadrupole moment $Q_m$ and
rotating with a pattern speed $\Omega_p$. The central concentration of
this model is characterised by its central density $\rho_c$. Model D
has a Dehnen-like bar potential \citep{Dehnen00} characterised by a
strength parameter ($\epsilon$) and a scale length ($\alpha$). The BW
model has a bar with a Barbanis \& Woltjer type potential
\citep{Barbanis.Woltjer67},
characterised by a strength parameter ($\hat\epsilon$) and a
scale length ($r_1$). The two latter models have the same axisymmetric
part as model A. Ferrers bars have a realistic density
distribution, but their non-axisymmetric force beyond corotation
decreases strongly with radius. The other two potentials are ad-hoc,
i.e. are not associated with a particular density distribution, but
can have strong non-axisymmetric forces beyond corotation. As in
Papers II and III, we use them to represent forcings not only from
bars, but also from spirals, oval discs and triaxial haloes. A full
description of these potentials can be found in Appendix A of Paper
III. These bar potentials will be used here together with different
types of rotation curves. As in our previous papers, we
measure angles in the standard 
mathematical sense, i.e. starting from the positive $x$ axis and
increasing anticlockwise.
 
As shown in Papers II and III, the properties of the manifolds depend
strongly on the relative strength of the non-axisymmetric forcing at
and somewhat beyond corotation. We measure this with the help of the
quantity $Q_{t,L_1}$, which is the value of $Q_t$

\begin{equation}
Q_t (r) = (\partial \Phi (r, \theta) /\partial \theta)_{max}/(r\partial
\Phi_0/\partial r),
\label{eq:Qt}
\end{equation}

\noindent
at $r_L$, the radius of the Lagrangian point $L_1$ or $L_2$, i.e. 
$Q_{t,L_1}=Q_t (r=r_L)$. This is not the same as
the strength of the bar, which is often measured by $Q_b$, i.e. the maximum
of $Q_t$ over all radii shorter than the bar extent. 
The radius at which this maximum occurs can be small compared
to $r_L$. So $Q_b$ is not necessarily a good proxy for 
$Q_{t,L1}$ and vice versa. Nevertheless, for the sake of brevity,
we will often replace in our discussions ``non-axisymmetric
forcings which are relatively strong at and beyond corotation'' simply
by ``strong bars'', or ``strong non-axisymmetric forcings''. 

\section{Morphology}
\label{sec:morpho}

Our aim in this work is to show that the dynamics of the manifolds
emanating from the unstable Lagrangian points can explain the
formation of spirals as well as of inner and outer rings in barred disc
galaxies. Thus, in this section, we will make the link between the
morphology of theoretical  
and of observed spirals and rings. At this stage, it is premature to make
a detailed comparison for one specific galaxy. Instead, we will make
global comparisons and make sure that there are no major
differences between models and observations and no obvious
discrepancies which would disqualify our theory.

\begin{figure*}
\centering
\includegraphics[scale=0.1754,angle=0.0]{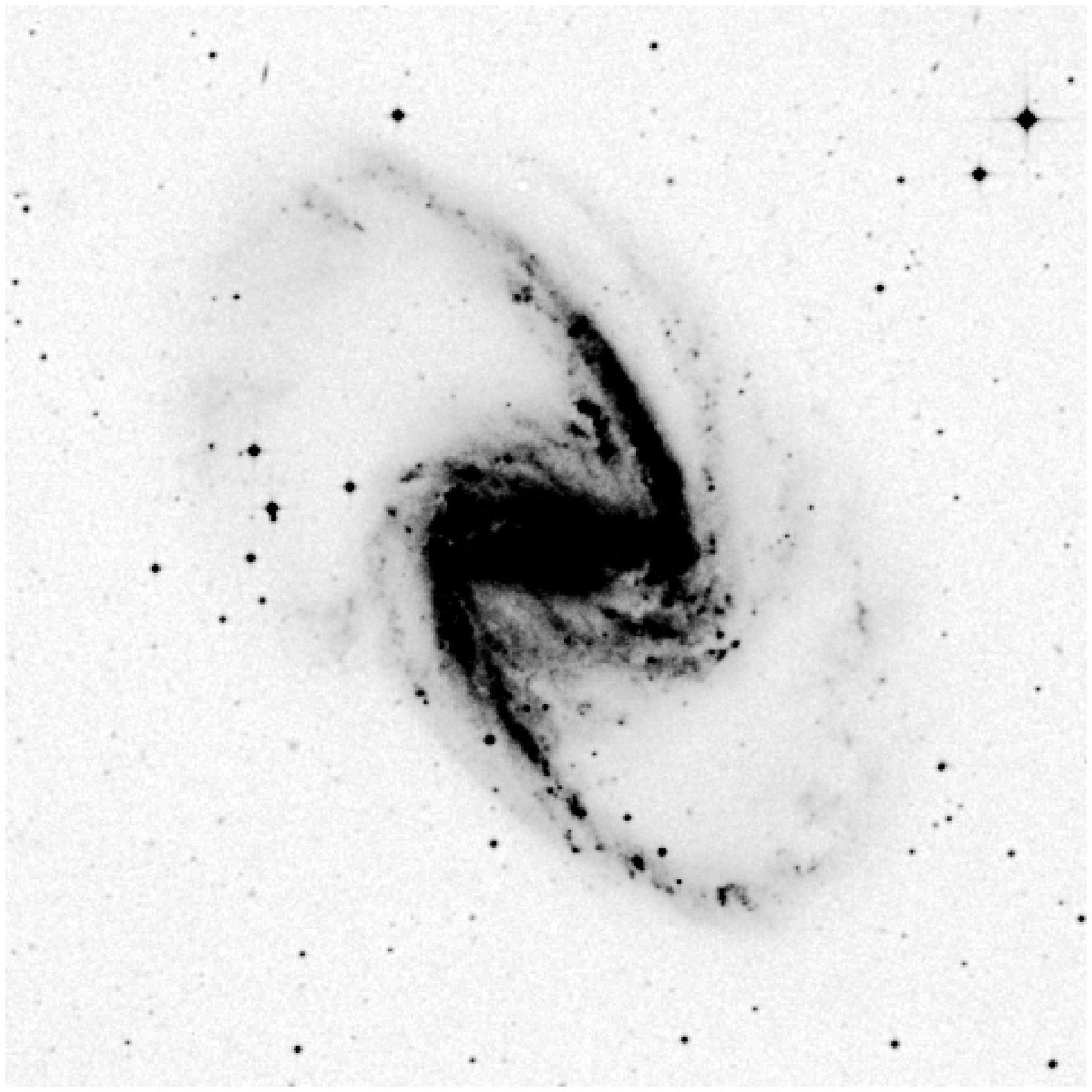} \hspace{0.2cm}
\includegraphics[scale=0.1686,angle=0.0]{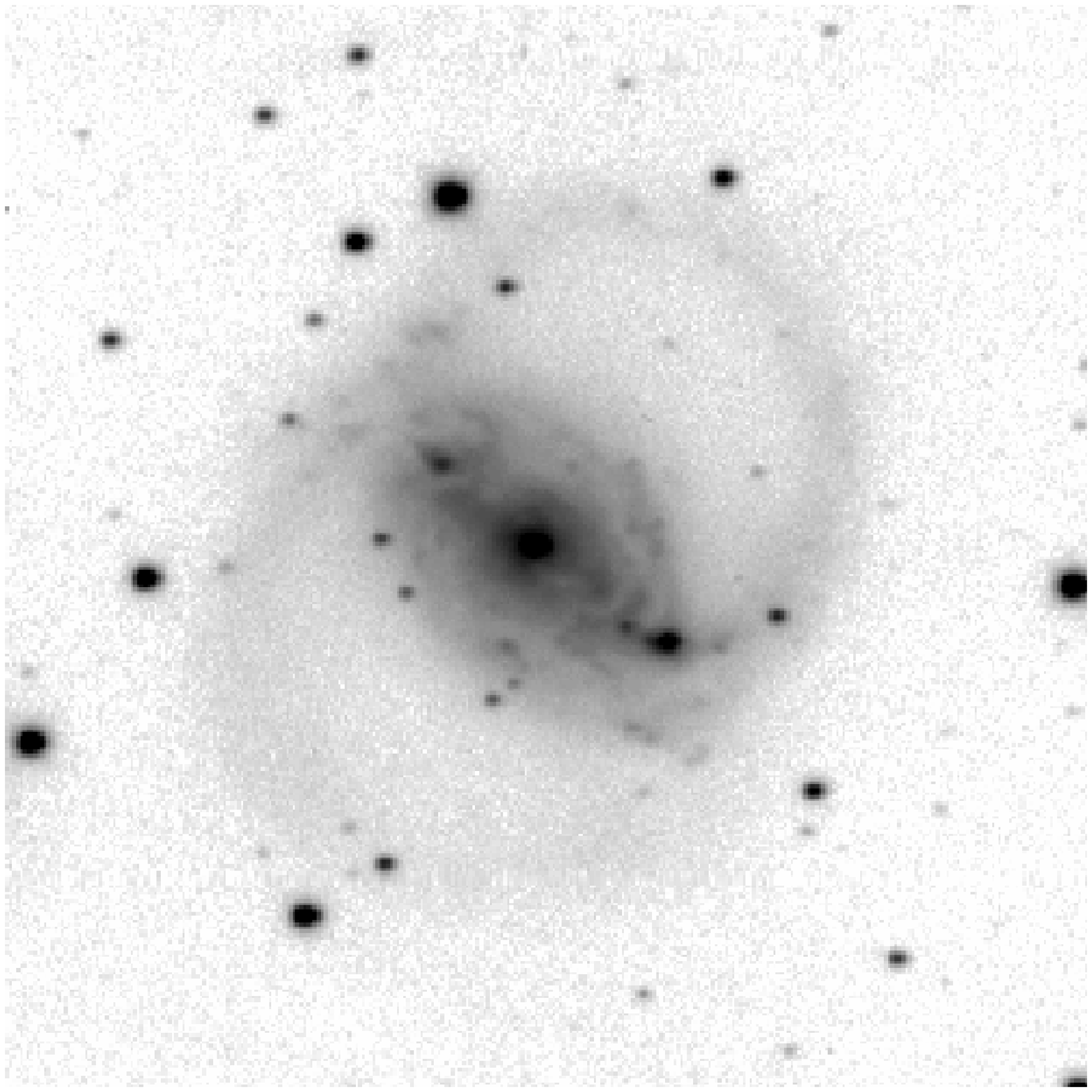} \hspace{0.2cm}
\includegraphics[scale=0.13294,angle=0.0]{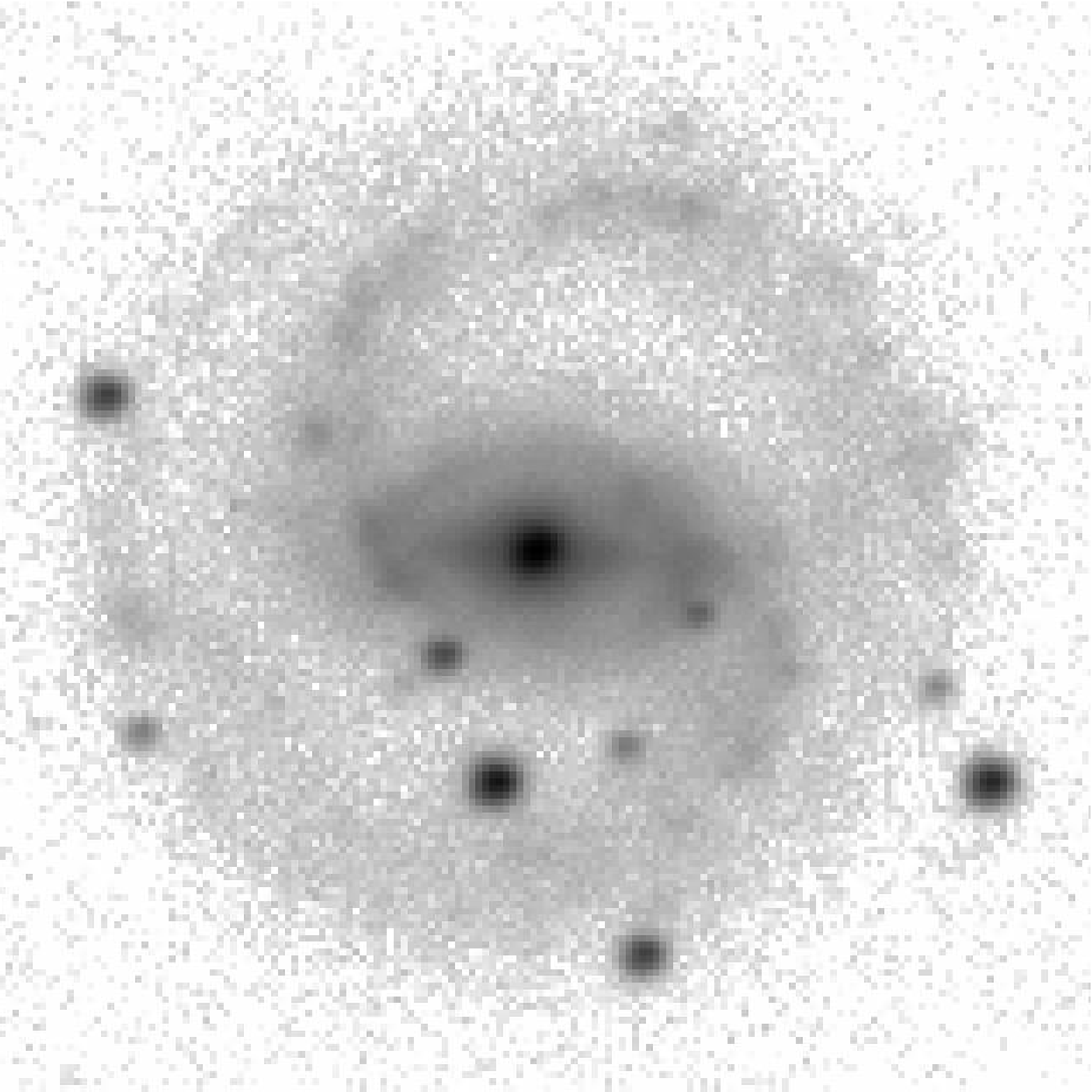}\hspace{0.2cm}
\includegraphics[scale=0.21,angle=0.0]{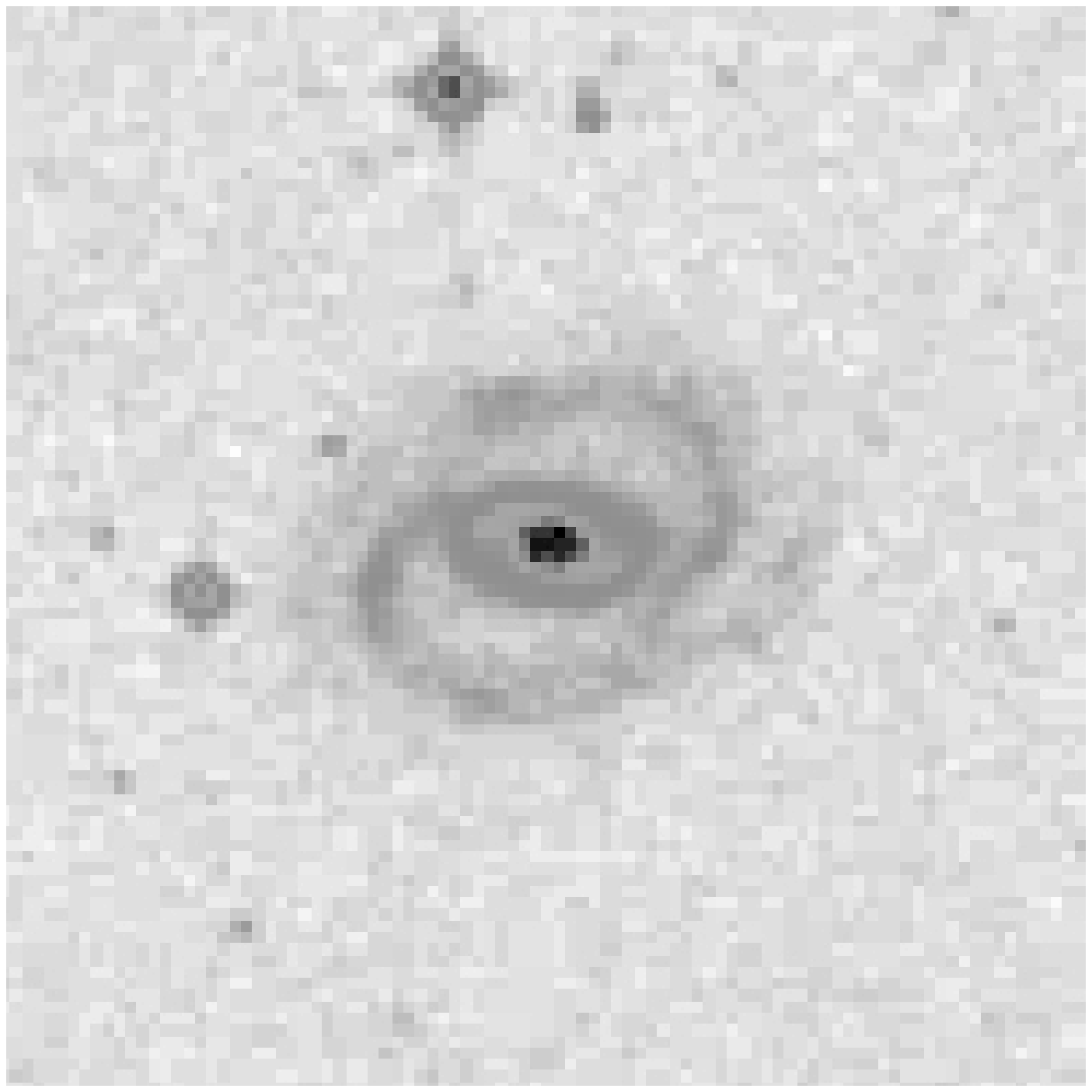}\\
\includegraphics[scale=0.3,angle=0.0]{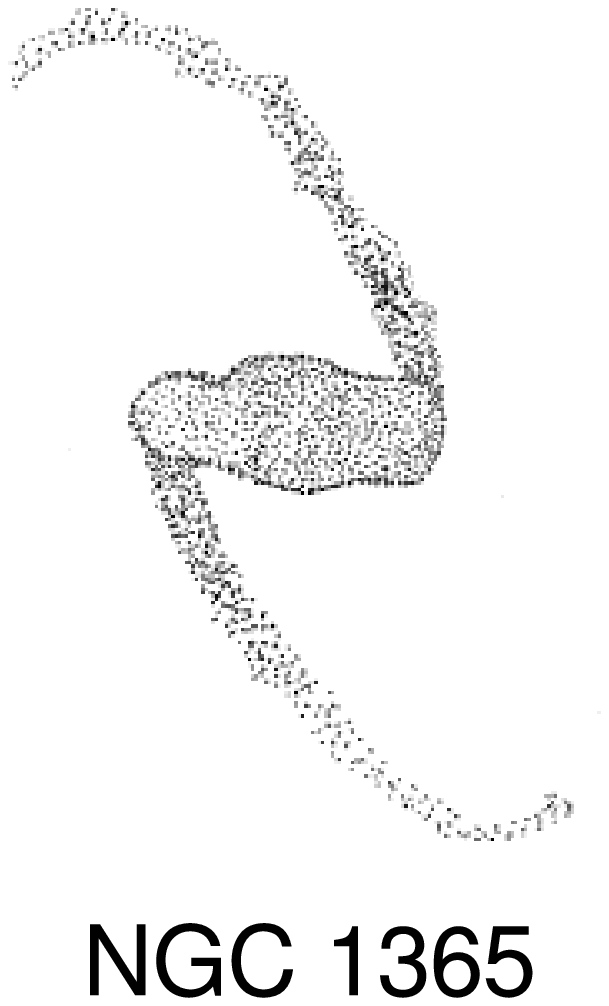}
\includegraphics[scale=0.7,angle=0.0]{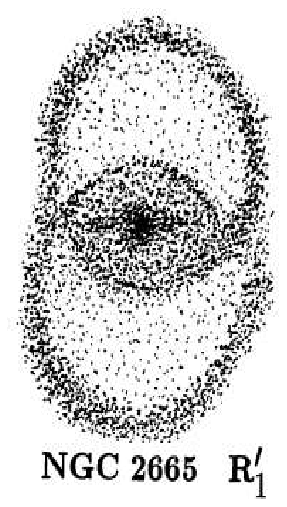}
\includegraphics[scale=0.7,angle=0.0]{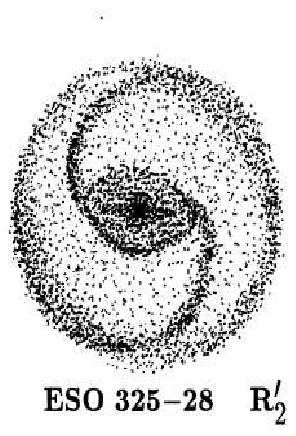}
\includegraphics[scale=0.7,angle=0.0]{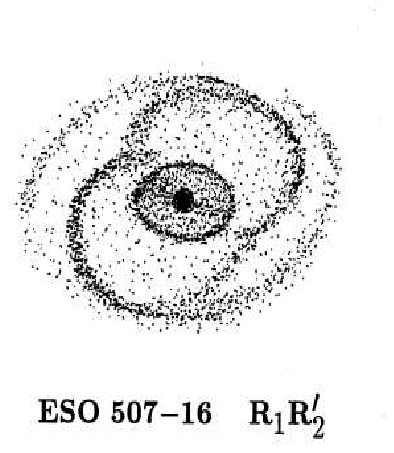}\\
\includegraphics[scale=0.197,angle=-90.0]{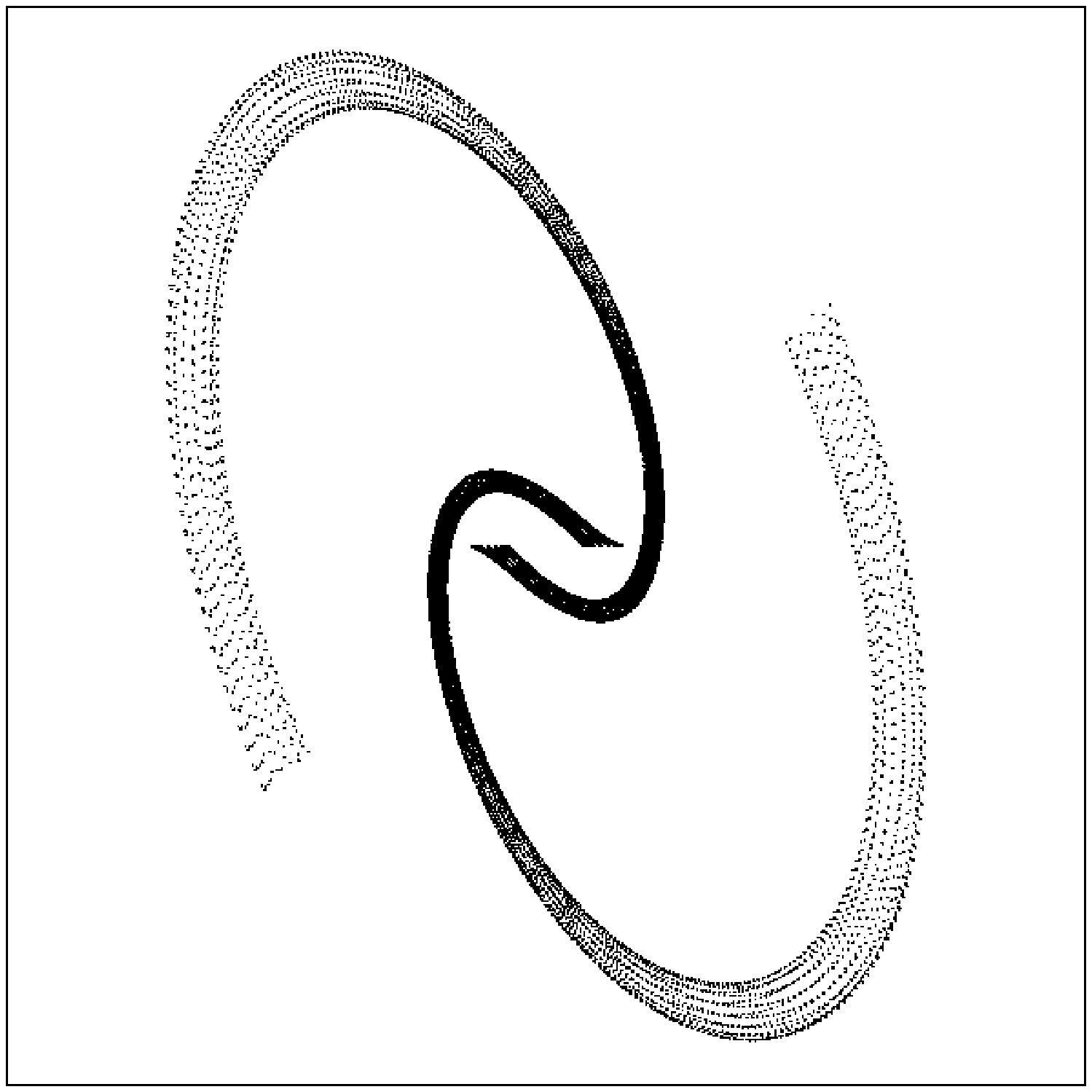}
\includegraphics[scale=0.197,angle=-90.0]{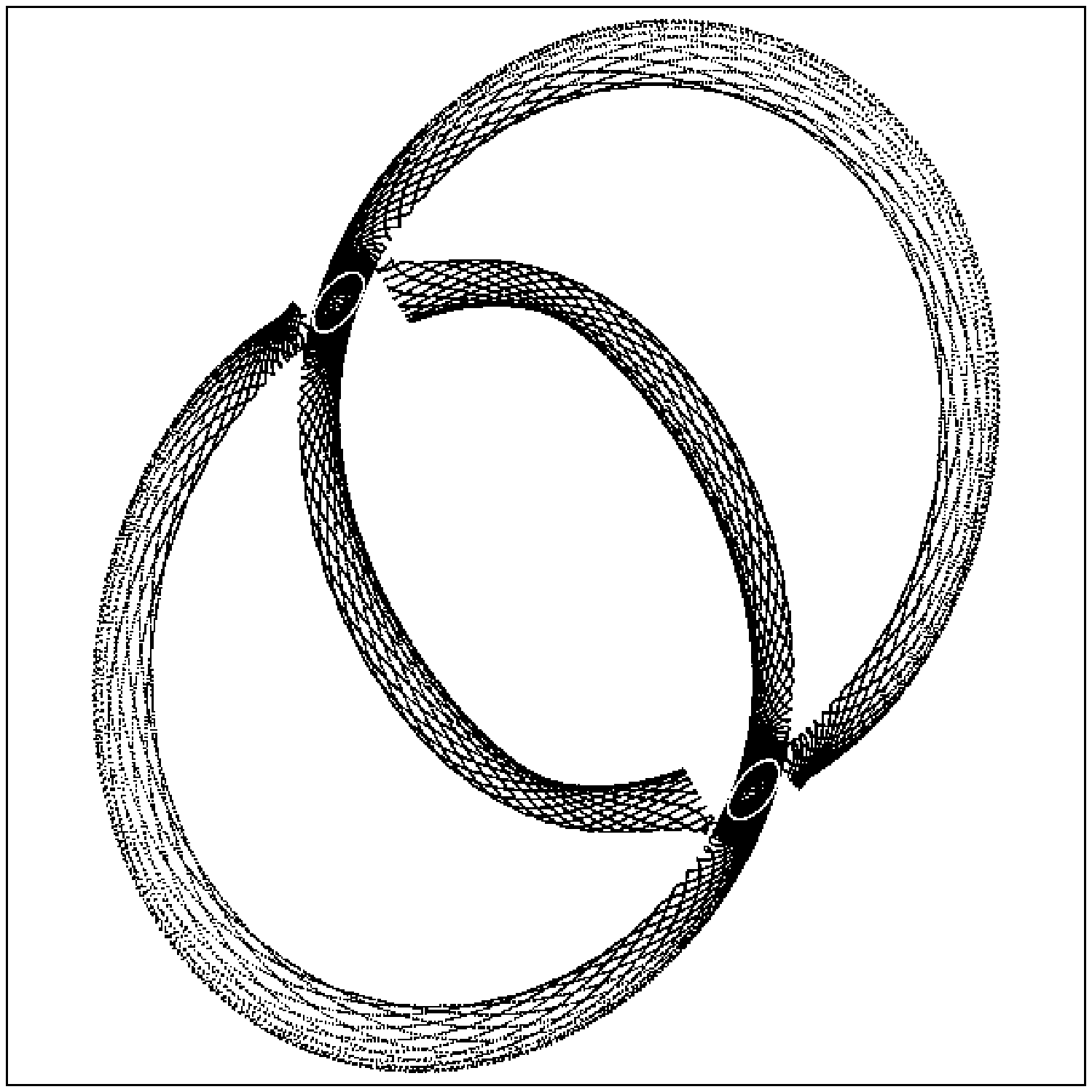}
\includegraphics[scale=0.197,angle=-90.0]{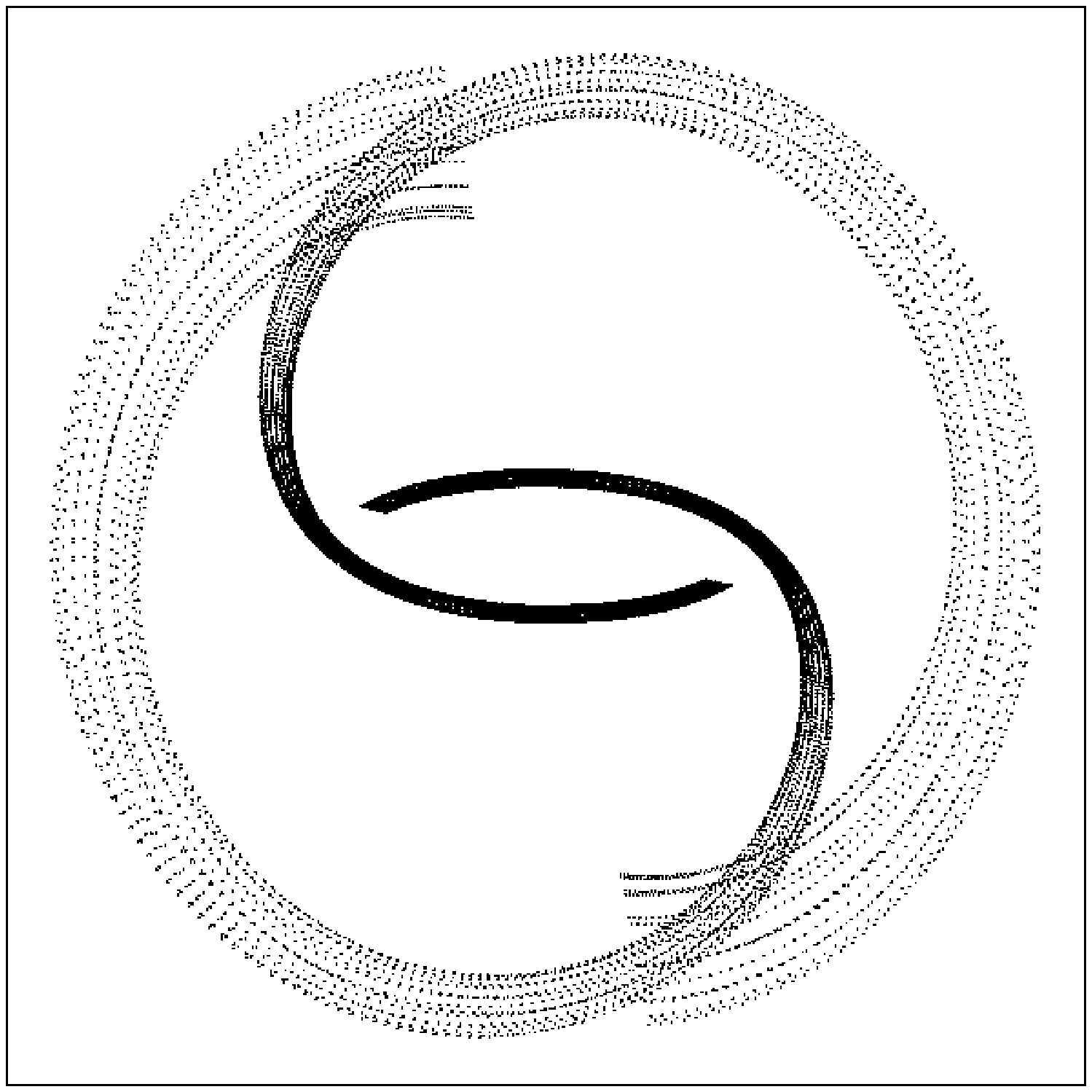}
\includegraphics[scale=0.197,angle=-90.0]{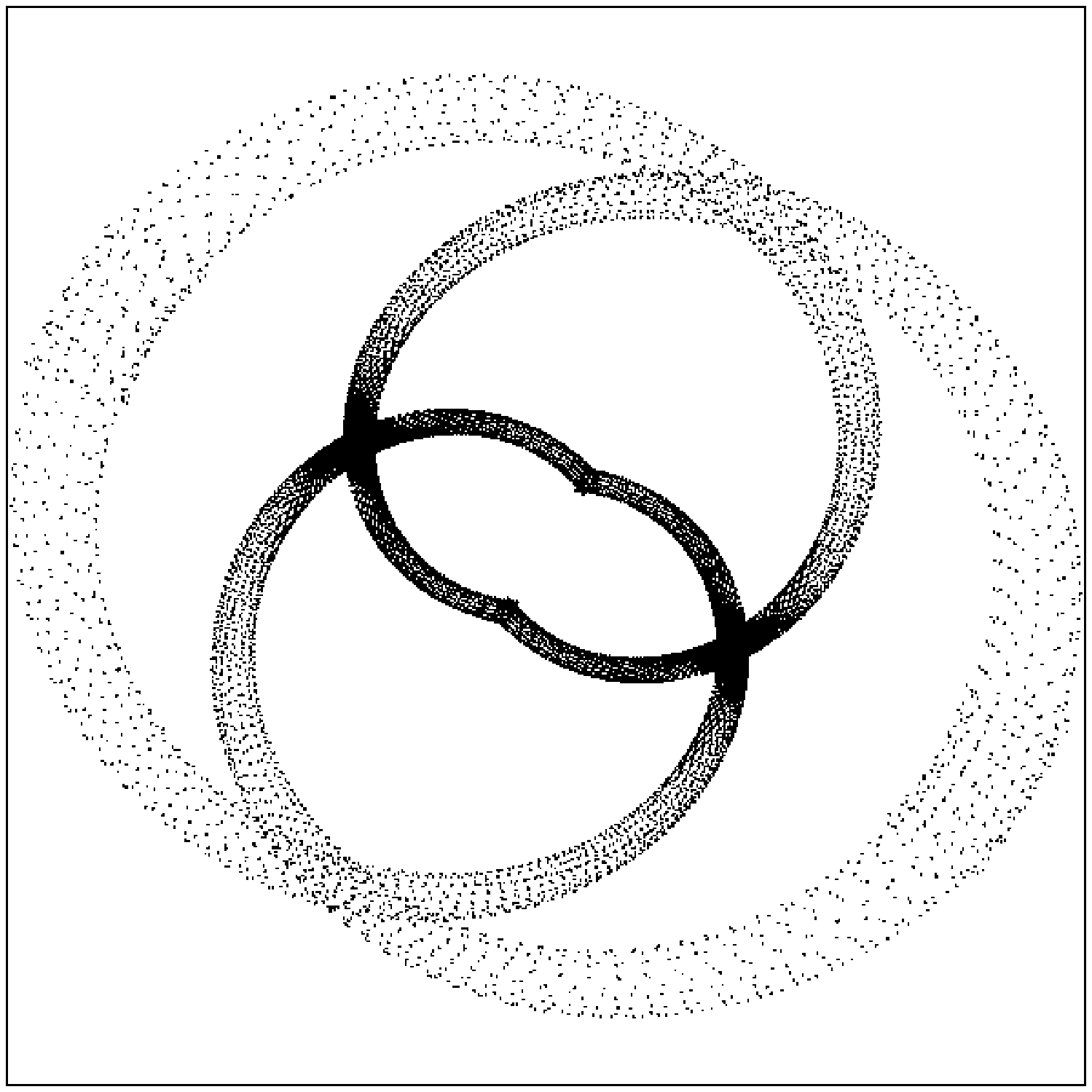}\\
\includegraphics[scale=0.22,angle=-90.0]{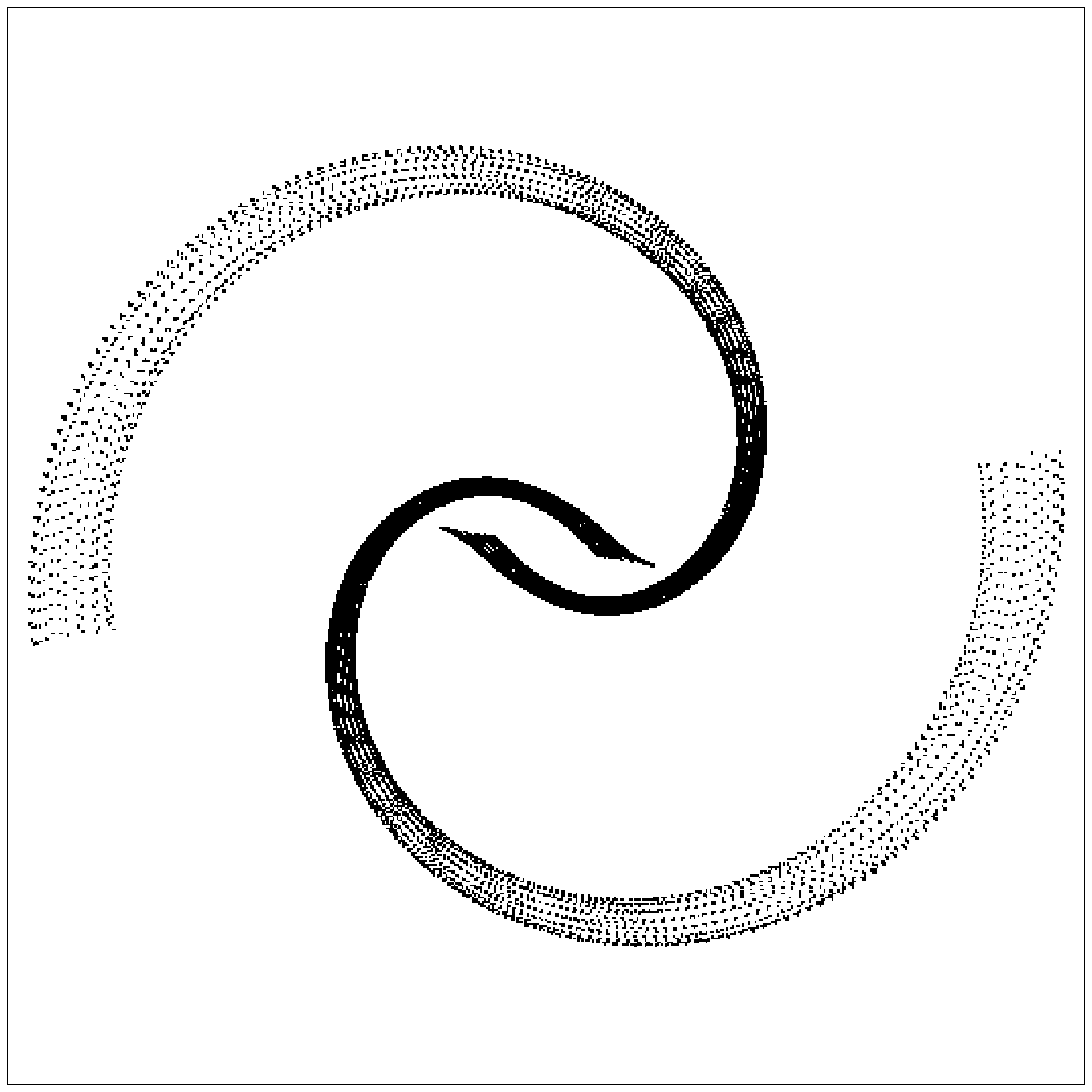}
\includegraphics[scale=0.22,angle=-90.0]{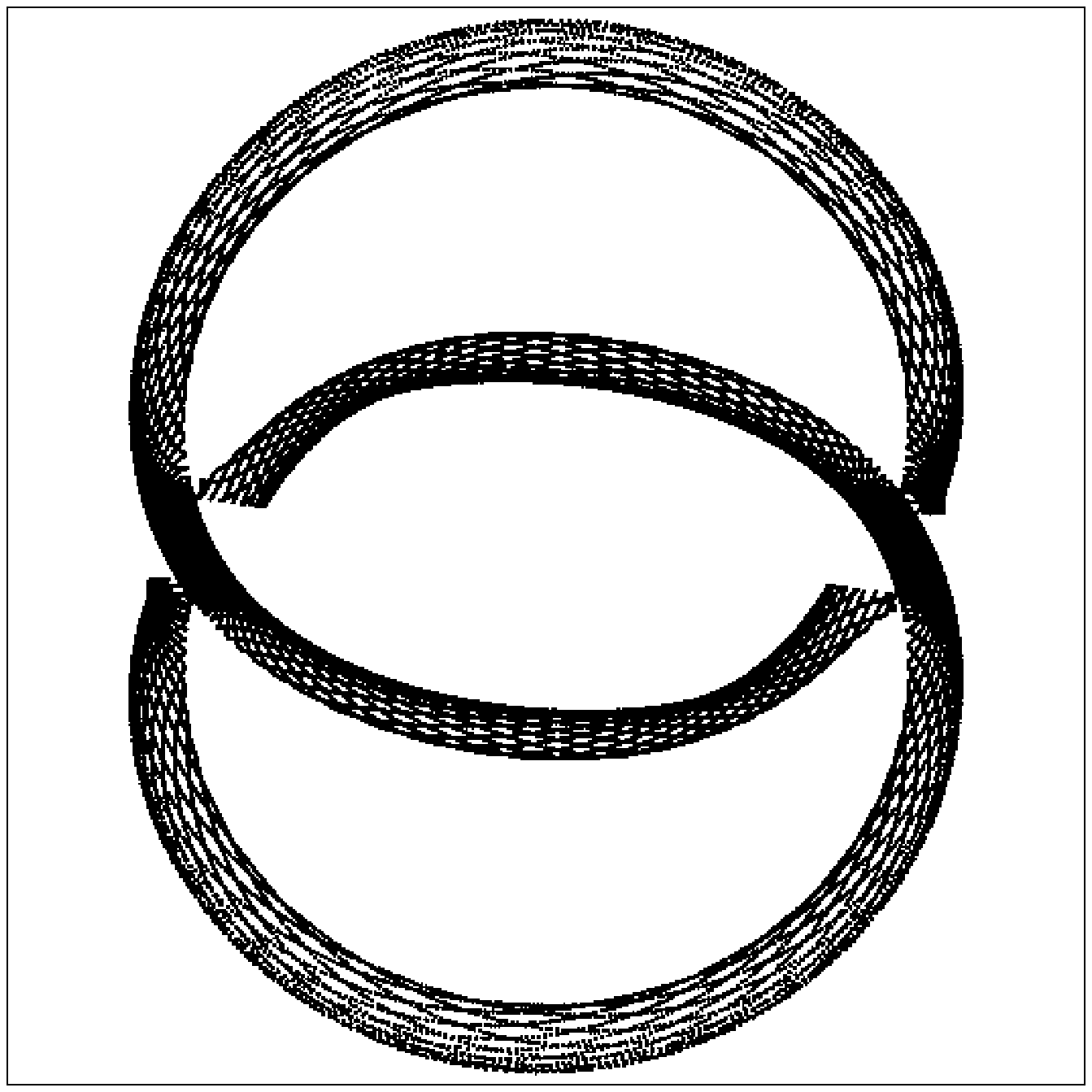}
\includegraphics[scale=0.22,angle=-90.0]{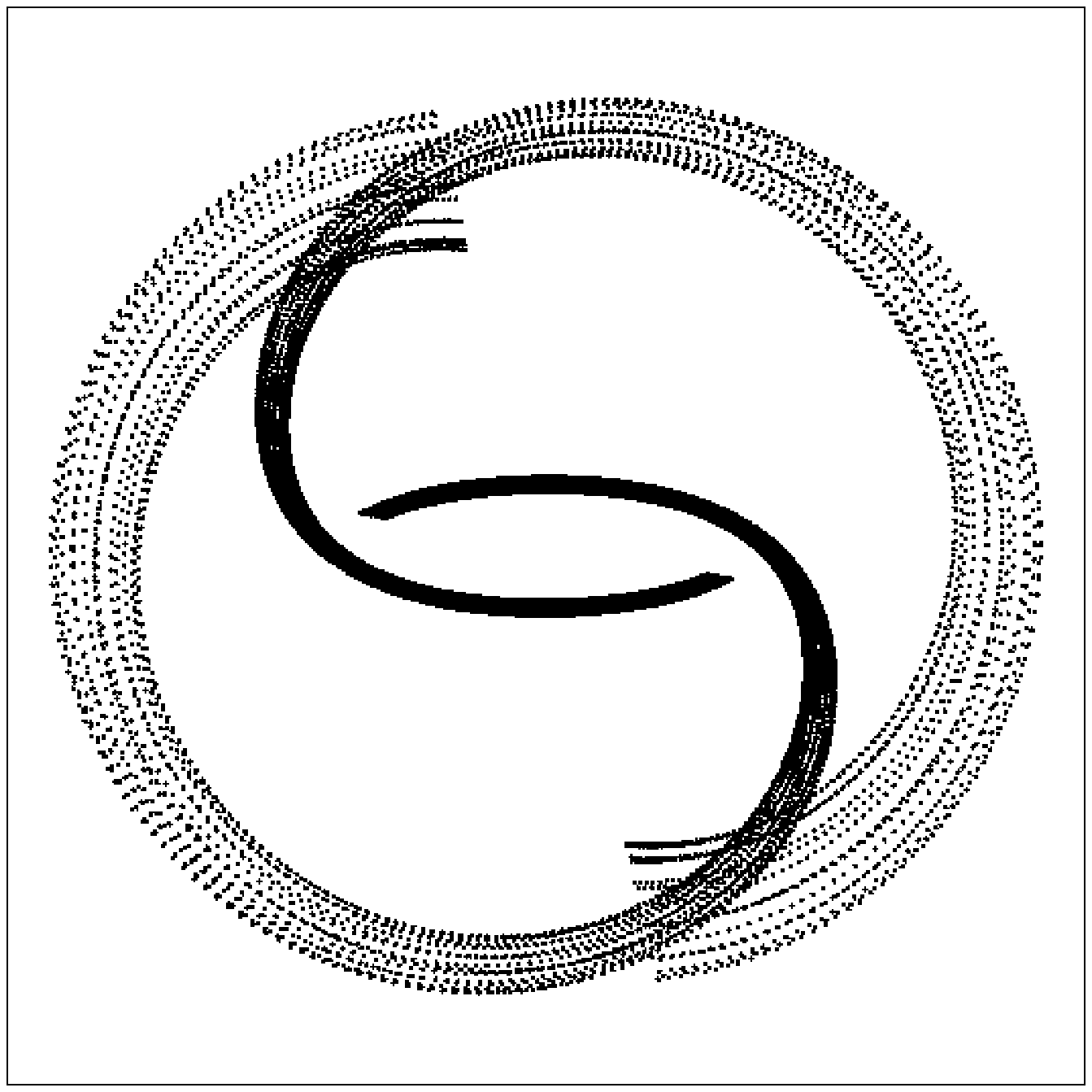}
\includegraphics[scale=0.187,angle=-90.0]{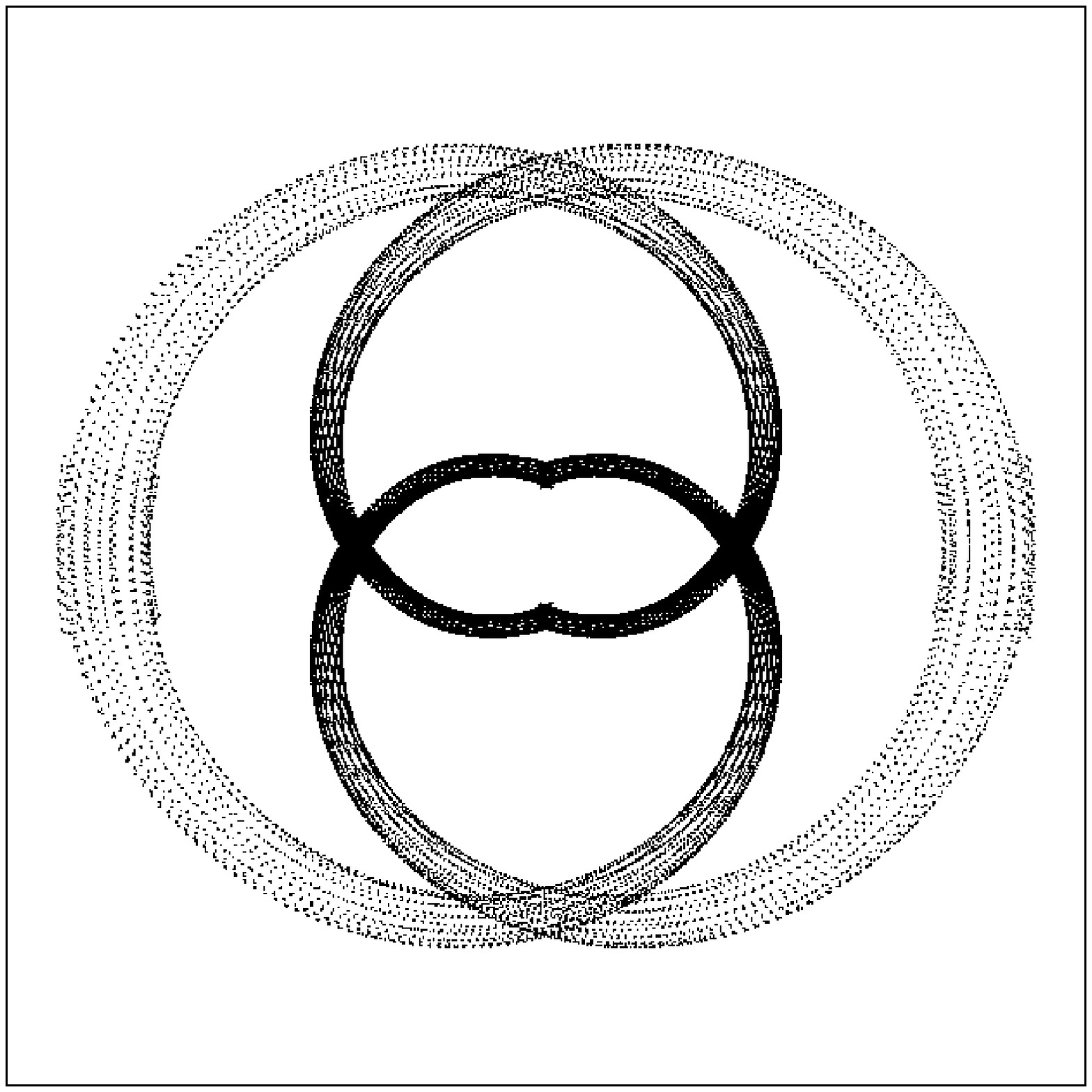}
\caption{Ring and spiral arm morphologies. The upper row shows four
galaxies, typical of the spiral, $R_1$, $R_2$ and $R_1R_2$
morphologies, namely NGC 1365, NGC 2665, ESO 325-28 and ESO 507-16,
respectively. The second row shows schematic plots of these galaxies,
bringing out the spiral and ring structures. The third row gives 
examples from our manifold calculations which have a similar morphology,
projected in roughly the same way as the observed galaxy or as the
schematic plot. Finally, the fourth row gives the face-on view of
these manifolds with the bar along the $x$ axis. Note that, contrary
to all other cases, the sense of rotation in the lowest left panel is
clockwise in order to ease the comparison with NGC 1365. 
}
\label{fig:family}
\end{figure*}

Any candidate theory should
reproduce correctly the observed spiral and ring morphologies. Our
theory can indeed accomplish this, as we will show in the next
subsections. The top two rows of panels of Fig.~\ref{fig:family}
\footnote{The images of
NGC 1365 (first column, first row) and of ESO 507-16 (third column,
first row) are taken from the Digital Sky Survey and those of NGC 2665
(second column, first row) and ESO 325-28 (third column, first row)
were reproduced from http://bama.ua.edu/~rbuta/devatlas \citep[see
  also][]{ButaCO07}. 
The three schematic plots in the second  
to fourth column and second row are from Buta \& Crocker (1991) and are
reproduced by permission of the American Astronomical Society.} show
four galaxies that contain archetypes examples of the 
characteristic morphologies that have to be reproduced; namely
spirals, inner rings and the three 
types of outer rings, $R_1$, $R_2$ and $R_1R_2$. They will be
described in detail in the following subsections. The upper
row of panels gives the images and the second row the schematic plots,
bringing out the characteristic morphologies of the galaxies, which the
manifolds have to reproduce
The third and fourth rows give manifold
examples which reproduce qualitatively the main characteristic 
features well -- the former viewed from an orientation similar
to that of the observed galaxy in order to permit a better
comparison, the latter viewed face-on.

Note that we have not used the potential of each specific galaxy, nor
have we made a 
search for the best fitting manifold. We simply found, amongst the
very large number of models for which we have calculated manifolds,
one which roughly represents the desired morphology.

\subsection{Spiral structure}
\label{subsec:spiralmorpho}

The left column of Fig.~\ref{fig:family} corresponds to 
NGC 1365, a  galaxy with a strong bar and a clear, well-developed,
grand design spiral structure. It has two spiral arms, each emanating
from one of the two ends of the bar and extending well into the outer
disc. The lower two 
panels in this column show a model qualitatively reproducing this
morphology. It also has two spiral arms, one emanating from each end
of the bar and extending outwards. The spiral winding also appears
similar and will be discussed in more detail in
Sect.~\ref{subsec:pitchangle}. Thus the model matches qualitatively
the morphological properties of the spiral very well.  

A further point to note is that in the case of escaping trajectories
-- i.e. trajectories that wind from $L_1$ (or $L_2$) outwards and do not
end at the same or the other Lagrangian point -- the
manifolds become wider as the distance from the centre increases (see
e.g. bottom left panel of Fig.~\ref{fig:family}). Under the assumption
that no mass is created or destroyed, this implies that the density in
this arm component should decrease with increasing distance from the
centre. However, if other arm components -- e.g. a weak,
  leading component --
are present, this will lead to bumps on an otherwise smoothly
decreasing density profile.

\subsection{Inner rings}
\label{subsec:iringmorpho}

Inner rings have roughly the same size as the bar \citep{Buta86}. 
\cite{Buta95} studied their orientation with respect to the bar major
axis using his catalogue of southern ring galaxies (CSRG). Comparing
with model distributions of apparent bar/ring position angles, he found
that inner rings are preferentially elongated in the direction of the
bar major axis.

In our models, some of the inner manifolds give rise to inner rings
while others contribute to the bar, particularly its outer parts. If
the $L_1$ and $L_2$ are not too strongly unstable, the inner
manifolds retrace the same path after one revolution around the
galactic centre (Paper III). In this case it is easy to get the outline of the
ring. A good example can be seen in the lower two rows of the second
column of panels in Fig.~\ref{fig:family}. Other examples can be found
in Papers I, II and III. In these cases, the inner manifold
branches outline a ring which has the same morphology as the observed
rings. Namely, it has roughly the same extent as the bar and is
elongated parallel to the bar major axis. In cases with strongly
unstable Lagrangian points, the inner manifold branches retrace the
same path only after a few revolutions. Thus the shape of the
corresponding density enhancement should be calculated by the
self-consistent superposition of the inner manifold loci, and this is
well beyond the scope of this paper. To summarise, the inner manifold
branches either reproduce well the shape and location of observed
inner rings, or populate mainly the outer parts of the bar.

\subsection{Outer rings}
\label{subsec:oringmorpho}

Outer rings are considerably larger than inner ones, with a major axis
of the order of twice the bar size. Depending on the relative orientation of
the major axes of the outer ring and of the bar, \citeauthor{Buta95}
(\citeyear{Buta95}) distinguished three different types of outer
rings. Rings of $R_1$ type are elongated perpendicular
to the bar major axis. Rings of
$R_2$ type are elongated along it and $R_1R_2$ have a part elongated
along the bar and a part elongated perpendicular to it. 

NGC 2665, ESO 325-28 and ESO 507-16, shown in Fig.~\ref{fig:family} (second
to fourth column of the top row), are prime examples of these
three types of outer rings. Note that NGC 2665 includes a
characteristic feature of $R_1$ rings. Namely at or near the direction
of the bar major axis, the ring approaches the 
ends of the bar. The features thus formed are called dimples
\citep{Buta86, Buta.Crocker91, Buta95} and give to these outer rings a
characteristic shape reminiscent of the number 8. 
Their schematic plots are given in the second
row. The third and fourth rows show examples of manifolds with similar
morphological characteristics. 

The example given in the bottom two rows of the second column reproduces
well the characteristic morphology of $rR_1$ rings in general and of
NGC 2665 in particular. It has both an inner and an outer
ring. The latter is
elongated perpendicular to the major axis and has the characteristic
morphology of a figure `8', with clear dimples near the two ends of
the bar. Thus, the manifolds reproduce well all the main observed
morphological features of observed $rR_1$ rings (see also Paper I).  

The example in the third column has a different morphological type;
that of an $R_2$ ring. There is a spiral starting from the end of
the bar, which, after winding by roughly 90$^\circ$, joins
with an outer ring. The latter is elongated along
the direction of the bar major axis, but its ellipticity is rather
small, i.e. it is not too far from circular. Also it does not have the 
dimples seen in the $R_1$ rings. These morphological characteristics
are shared by both the galaxy and the manifold.

The example in the fourth column is of the $R_1R_2$ morphology and 
has two outer structures, i.e. two 
outer rings. The inside one is elongated perpendicular to the bar
major axis and has characteristics dimples near the ends of
the bar, i.e. it is of the $R_1$ variety. The outermost one is
elongated along the bar major axis, has a relatively small elongation
and no dimples in the direction of the bar major axis. 
All these morphological characteristics are clearly seen in the
manifold examples in the last two rows. 
Also, in all the examples of $R_1R_2$ morphology we have calculated so
far, the $R_1$ feature lies within the $R_2$ feature and this also
holds for all the observed cases we have seen.  

To summarise, we can say that the main morphological features of the three
types of outer rings are qualitatively very well reproduced by our
theoretical results. 

\subsection{Bar strength and morphological types}
\label{subsec:strengthmorpho}

In Paper III we showed that the morphological type -- i.e. whether we
have a spiral, an $R_1$, an $R_2$, or an $R_1R_2$ ring or pseudo-ring
-- depends on 
the strength of the non-axisymmetric forcing at the corotation
radius. Relatively weak non-axisymmetric forcings give $R_1$ and
$R_1'$ type morphologies, while stronger forcings give spirals and the
remaining types of rings.

A related result was found in Paper II, where we showed 
that the morphology can also depend on whether the rotation curve is
rising, flat or slightly falling in the relevant part of the
disc, i.e. at and somewhat beyond the Lagrangian radius. In examples
given in Fig. 8 of that paper, we showed that, for the same bar, a
rising rotation curve can give a $R_1$ or $R_1'$ morphology, while a
decreasing rotation will give an $R_2$ or a spiral morphology. This
can now be understood from the results presented 
in Paper III. Indeed a different axisymmetric model will change the
values of $Q_t$ at and beyond the Lagrangian radius, and therefore the
morphology and the properties of the rings and spirals. 
Rising rotation curves imply more mass in the relevant part
of the galaxy and therefore a stronger axisymmetric force and a smaller
$Q_{t,L_1}$ value, the opposite being true for declining rotation
curves. Thus the result found in Paper II is in fact a particular
aspect of what was found in Paper III, namely that the morphological
type depends on $Q_{t,L_1}$, the value of $Q_t$ at and somewhat beyond
the Lagrangian radius.  

This is a clear prediction, which can be tested by observations. In
doing so, two points need to be kept in mind. The first one is that
the relative importance of the non-axisymmetric forcing depends not
only on the bar or spiral, but also on whether the rotation curve is
flat, rising or somewhat falling (Paper II). Thus such a test is not
possible solely with photometric data, but necessitates also
kinematics. The second point to be
kept in mind is that a spiral forcing can produce spirals easier than
a bar forcing of comparable amplitude. Thus, when comparing
spiral to ring morphologies one should keep in mind that the potential
of the spiral has to be also taken into account. This has been
already argued, albeit in a different context, by \cite{Lindblad.LA96}. 

Making a quantitative comparison is, at this stage, not
straightforward because our theory is not self-consistent and because
we only use a bar forcing, while in real galaxies the
rings and spirals are formed under the combined
effect of the bar and of their own potential. The bar forcing is of
course much stronger than the spiral one. \cite{ButaVSL05} give for
the SB galaxies in their sample for the bar an average $Q_B$ of 0.294
$\pm$ 0.21 and for the spiral an average $Q_S$ of 0.171 $\pm$
0.014. Similarly, \cite{DurbalaBSVM09}, using a sample of isolated
barred galaxies of type Sb-Sc, find 0.243 $\pm$ 0.015 and 0.166 $\pm$ 0.009
for the mean values of $Q_B$ and $Q_S$, respectively. For the medians
they find 0.222 and 0.165, respectively. These numbers can not be
directly translated into a ratio of 
relative effects, because the maximum of the spiral forcing occurs at
larger radii than that of the bar forcing. They nevertheless clearly
show that the effect of self-consistency will change somewhat the limits
between the zones with different morphologies ($R_1$, $R_1'$,
spirals, $R_2$, $R_1R_2$ etc; see Paper III), shifting them to lower bar
strengths. The amplitude of this shift 
will depend on the strength of the spiral, so that no quantitative
comparison is possible at this stage. 

\cite{ButaVSL05} measured the bar strength for 147 spiral galaxies in
the statistically well-defined Ohio State University Bright Galaxy
Survey (OSUBGS; \cite{Eskridge+00}) sample and found that the average bar
strength in barred galaxies with spirals (SB(s) types) is $<Q_b>=0.329$,
i.e. higher than the average for all barred galaxies, which
is  $<Q_b>=0.294$. They also find that the average bar
strength in barred galaxies with inner rings (SB(r) types) is $<Q_b>=0.201$,
i.e. lower  than the average for all barred galaxies. These
trends are rough, but still are in good agreement with our theoretical
predictions from Paper III. 

\section{Main spiral properties}
\label{sec:spiralprop}

\subsection{Why are spirals trailing?}
\label{subsec:trailing}

Simply by viewing a spiral galaxy projected on the sky, it is not
possible to say whether the arms are trailing or leading, even if one
knows the sense of rotation of the material within the galaxy
disc. As explained by \cite{Binney.Tremaine08} and
\cite{Binney.Merrifield98}, one needs also to know
which side of the galaxy is nearest to us. This can be obtained by
studying the obscuration by dust (see e.g. \citealt{deVaucouleurs59}
and \citealt{Sandage61} for such pioneering studies). In almost all
cases where the result is unambiguous the arms are found to be
trailing. In our own Galaxy also the arms are trailing. It is thus
generally admitted that spiral arms in disc galaxies are trailing. How
does that compare with our theoretical predictions?

The anti-spiral theorem \citep{Lynden-Bell.Ostriker67} states that if a
steady-state solution of a time-reversible set of equations has the
form of a trailing spiral, then there must be an identical solution in
the form of a leading spiral. Thus, density wave theory can not
distinguish between trailing and 
leading solutions, unless one takes into account the growth of the
spiral. 

Swing amplified spirals, however, are not stationary,
so do not have this shortcoming \citep{Toomre81}. The inwards going
wave is leading and will be reflected at the centre of the galaxy into an
outwards-going trailing wave which will be amplified in the corotation
region. Therefore, according to this theory, both a
trailing and a leading component will be present, but since a very
strong amplification occurs, the trailing component will have a much
larger amplitude than the leading component. Thus, visually one will
see a trailing spiral, whose amplitude will present oscillations with
radius due to the interaction of the leading and trailing spiral
phases \citep{Atha84, Atha92b, PuerariBEFE00}. 

Since the potentials we have used so far are stationary in the
rotating frame of reference, the anti-spiral
theorem will apply. Indeed, if one reverses the arrow of time in the case of
the unstable manifolds -- which have a trailing spiral shape -- one finds the
stable manifolds -- which are of the same shape but leading. This is true
whenever the bar does not evolve. Material, however, needs initially
to get trapped into the manifolds, and this can only be done while the
bar is being formed, 
or while it grows. This introduces time evolution and thus breaks the
degeneracy between trailing and leading arms. 
To understand this further, we made
a number of response simulations. For this we considered 10\,000 
particles initially in circular orbits around the galactic
centre and grew gradually (in one bar rotation) a bar in the disc.

Let us start by describing the results of a case which, when the bar
has reached its final amplitude, will have a spiral manifold
morphology. The response is a trailing spiral which forms gradually
while the bar grows. This means that at any time there are more
particles in the trailing arm than in the background or the
corresponding symmetric leading arm. This, however, does not, on its
own, prove that the unstable (trailing) manifolds trap particles as
the bar grows, because particles could still traverse the arm, which
would then be a density wave. 

To reveal the differences between the unstable and stable manifold
and to prove that it is mainly the unstable manifold that traps particles,
we need to work in phase space rather than configuration space,
i.e. we need to consider velocities as well as positions.
We thus chose a number of simulation particles at
random in the radial range of the spiral arms
and in the energy range where the stable and unstable
manifolds exist. For each of these, we calculated the manifolds at the
particle energy and found the tube they outline in phase space. We 
then checked 
whether the particle is trapped by the stable or the unstable
manifold, or whether it is not trapped at all. We repeated this for
several time steps and a large number of particles and from these
results we can get some statistical information. We find that of the
order of 30 -- 50\% of the particles are trapped by the unstable
manifold and only 0 -- 2\% by the stable ones. The remaining particles
are either not trapped by any manifold, or it was not possible to
give an unambiguous answer due to their locations (e.g. close to the
Lagrangian points). This very important numerical difference between
the stable and the unstable manifolds shows that our
theory can make a strong prediction concerning the sense of winding of
the spiral arms. Namely, it predicts that as the bar grows in a
galaxy, the unstable manifolds will trap mass elements. Since the
unstable manifolds have a winding in the trailing sense, this means
that our theory predicts that spiral arms generated by the manifolds
of the $L_1$ and the $l_2$ Lagrangian points should be trailing, 
as observed.

The situation is different in the case of a bar which, after
the bar has reached its maximum amplitude, has manifolds with
the shape of rings. As we already discussed in Paper II, in this case
the unstable and stable manifolds connect in phase space, so it is not
possible to say whether a particle belongs to the stable or the
unstable manifold. Thus, in this case, both stable and unstable
manifolds can be populated, so that there is no preference of trailing
or of leading windings. This is in good agreement with the observed
shape of rings in real galaxies, which have no preferred winding. 

Thus to summarise, in the case of spirals, material gets trapped
mainly in the manifolds as the bar grows, mainly in the unstable
one. This accounts for the preponderance of trailing (as opposed to
leading) spirals. In the case of rings, the stable and unstable
manifolds communicate in phase space, so there is no preferred sense of
winding. 
 
\subsection{Number of arms}
\label{subsec:narms}

In the vast majority of cases, spirals in barred galaxies have two
spiral arms. This is in good agreement with our theory which 
relates the spiral arms to the unstable Lagrangian points. In the
standard case, a barred galaxy has two unstable Lagrangian points,
$L_1$ and $L_2$, which are saddle points (Paper I). Each one accounts for a
spiral arm. Thus our standard theory predicts that barred galaxies
will have two-armed grand design spirals
emanating from a location near one of the two opposite ends of the
bar. Since all the forcings we have used so far are bisymmetric, we
found by necessity two symmetric spirals. In real galaxies, however,
there are always asymmetries in the forcing potential, and this will
lead also to asymmetries 
between the two spirals, or to small angular shifts of their beginning with
respect to the tip of the bar. There is thus very good agreement
between the results of our theory and observed spirals regarding the
number of spiral arms. 

Although rare, there are, nevertheless, a few observed barred
galaxies with more arms, in particular with a four armed structure
\citep{Buta95}. A good example is ESO 566-24. Can our theory account 
for such structures?

If more than two Lagrangian points with saddle behaviour were present,
our theory could predict spirals of higher multiplicity. 
The most straightforward way of
achieving this is to include a $cos(m\theta)$ forcing, with $m >
2$ and a sufficiently high amplitude. Making a full analysis of such
potentials, of their origin and of the corresponding manifolds is
beyond the scope of this paper. We will, nevertheless, examine 
the iso-effective-potential curves and the type of stability of the
corresponding Lagrangian points for some specific cases, in order to
get some insight and to test whether such structures are possible. We
will also, for simplicity, limit ourselves to $m$ = 4 cases, but the
extension to other $m$ values is straightforward.

Ferrers bars have eightfold symmetry and forcing terms $m$ = 2 and
higher. Increasing the bar mass increases both the $m$ = 2 and the $m$
= 4 terms leaving their ratio unchanged. Increasing the bar ellipticity, 
however, changes the $m$ = 4 over $m$ = 2 ratio; thinner bars having
larger values of this ratio. In such bars the $L_4$ and $L_5$ can become
unstable \citep{Atha92a}, but they become complex unstable, not saddle
points. These do have manifolds, both stable and unstable, but it is
unclear whether these can account for spiral arms and at a first glance
it does not seem very promising. Further theoretical work is
necessary to clarify this issue.

\begin{figure}
\centering
\hspace{0.5cm}
\includegraphics[scale=0.4,angle=-90.0]{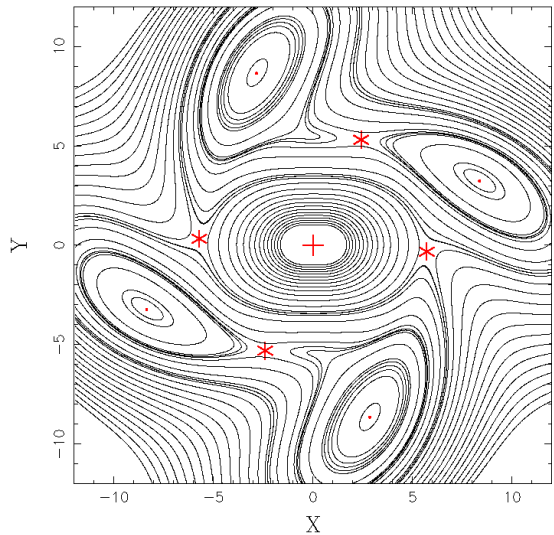} 
\caption{Iso-effective contours for a potential with an $m$ = 2
  symmetry inside corotation and an $m$ = 4 outside it. There are
  four Lagrangian points in the region around corotation and four more
  somewhat beyond that. The former are of saddle type and are
  marked with a red star. the latter are stable and are marked with a
  red dot. The four-armed barred galaxies, such as ESO 566-24, 
  could be due to potentials of similar type.
}
\label{fig:m4a}
\end{figure}

The simplest and most straightforward bar potential having the desired
properties is of the form

\begin{displaymath}
\Phi_{b}(r,\theta)=A_2(r)\cos(2\theta)+A_4(r) \cos(4\theta+\phi(r)+\theta_0),
\label{eq:m4a}
\end{displaymath}

\noindent
This is bar-like within corotation with only an $m$=2 forcing and
spiral-like with $m$ = 4 outside it. Fig.~\ref{fig:m4a} shows 
a specific example, whose iso-effective-potential plot looks
promising. Around corotation there are four unstable
Lagrangian points of the saddle type, similar to those at $L_1$ and $L_2$
in the standard case. At a somewhat larger radius there are four
stable Lagrangian points, similar to the $L_4$ and $L_5$ in the standard
case. Introducing such an $m$ forcing is not the only way to obtain
four unstable Lagrangian points with a saddle point behaviour. 
This can also be achieved with an $m$ = 2 only, provided it
has a non-monotonically decreasing profile of the appropriate shape. 
Again we stress that these are very simple ad hoc potentials, which may
not necessarily be linked to a realistic density distribution and do
not represent any galaxy in particular. It is nevertheless useful to
consider them, because they may give some insight as to what kind of
potential can explain  
the spiral structure in ESO 566-24 \citep{RautiainenSB04}, or other,
undoubtedly very rare, galaxies of this type.

To summarise, our theory predicts that the vast majority of  barred
spirals will be two armed, but that, for certain types of potentials,
it is possible to have more arms. This is in good agreement with what 
is indeed found by observations. 

\subsection{Spiral arm shapes}
\label{subsec:pitchangle}

\begin{figure}
\centering
\hspace{0.5cm}
\includegraphics[scale=0.33,angle=-90.0]{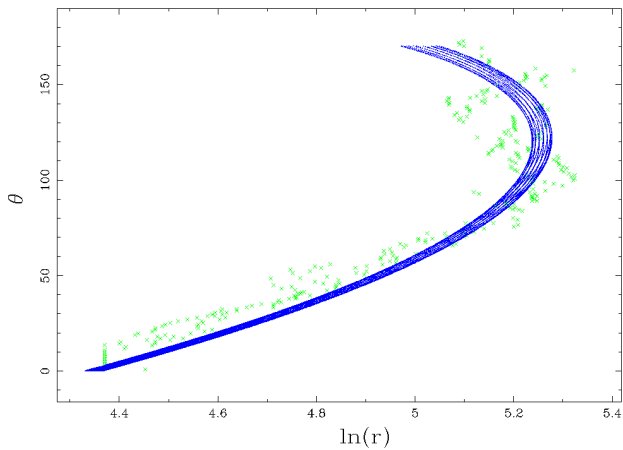} 
\caption{Comparison of the manifold loci on the $\theta = ln (r)$ 
  plane (blue on the online version, black in the paper version) with
  the spiral arm loci in NGC 1365 (x signs, green in the online
  version, gray in the paper version). 
}
\label{fig:pitchangle1365}
\end{figure}

We showed above that the manifolds have shapes that can reproduce
spiral arms. But  
how does the winding of the manifold loci compare with that of spiral
arms in real barred galaxies? To answer this question we deprojected
the image of NGC 1365 (upper two panels of the left column of
Fig.~\ref{fig:family}), using viewing angle values obtained by
rounding off the estimates given by \cite{JorsaterVan Moorsel95},
namely a position angle of 220$^\circ$ and an 
inclination angle of 50$^\circ$, and we measured at
each radius the coordinates of the maxima along the arm. We can now
compare the loci outlined by these
measurements to the loci of a typical spiral manifold. For this we
chose a D model with $\epsilon$ = 0.05 and $r_L$ = 5. In 
Fig.~\ref{fig:pitchangle1365} we can see that the comparison is
very satisfactory. In particular, both observed spirals and calculated
manifolds share the common property that after reaching a maximum
radius they fall back towards the inner parts of the galaxy. For a
more precise comparison we would need to 
calculate the potential of NGC 1365 from some image in the near
infrared as well as kinematics, but this is beyond the scope of this paper.

Thus our theory, explaining the formation of spiral arms in barred
galaxies with the help of the invariant manifolds of the $L_1$ and
$L_2$ Lagrangian points, explains the characteristic winding of the
spiral arms observed in a large fraction of barred
galaxies. Namely, as the angle increases the radius first increases
and, upon reaching a maximum radius, `falls back' towards the bar
region, or towards the region of the other spiral. This
property has not been so far reproduced with density wave models,
where the spiral arms tighten up 
as they reach outer Lindblad resonance, but do not fall back
(\citealt{Lin.YS69} and later works of this group, \citealt{Atha80} for
an approach not relying on the WKBJ approximation). 
 
\subsection{Trends with the pitch angle}
\label{subsec:pitchangletrends}

\begin{figure}
\centering
\hspace{0.5cm}
\includegraphics[scale=0.3,angle=-90.0]{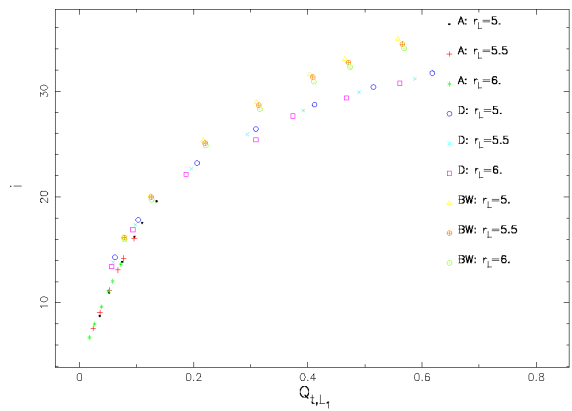} 
\caption{Pitch angle $i$ as a
  function of bar strength measured by $Q_{t,L_1}$. We plot the results
  only for three values of $r_l$ -- namely $r_L$ = 5, 
  $r_L$ = 5.5 and $r_L$ = 6 --
  but we have verified that other realistic
  values also show the same trend. We include in the plot models from
  all three families, A, D and BW. For A models we include in the plot
  only models with a Ferrers' index $n$ = 1, but we verified that
  models with $n$ = 0 lie on the same curve. 
}
\label{fig:allpitchf(Qb)}
\end{figure}

In order to assess how various bar parameters affect the winding
of the spiral, 
we obtained the value of the pitch angle by fitting a logarithmic
spiral
$$\theta - \theta_0 = cot~i~ ln (r/r_0)$$
over the main relevant part of the manifold, for many
models. We examined each of our three model families separately
(families A, D and BW), and used as a measure of the bar strength
$Q_m$, $\epsilon$ and $\hat{\epsilon}$, respectively. We find that (not
shown here), for all three families, the pitch angle increases with
increasing bar strength and the form of the dependence is similar. 
In other words, the spirals corresponding to stronger bars are more
open than those from weaker bars. This is a direct consequence of the
spiral properties described in section 3 of Paper III. 
To our knowledge, no such relation has as yet been reported by
observations. It would be thus very useful to test whether such a
relationship can indeed be found. In order for this to be possible, we
need first to check whether the trend persists when the pitch angle of the
various models is plotted as a function of a common bar strength
measure. For the latter we use $Q_{t,L_1}$ 
and show the results in Fig.~\ref{fig:allpitchf(Qb)} for the three
model families and three pattern speed values. The trend is
clearly present, with relatively little spread due to the different
models. The pitch angle increases steeply as $Q_{t,L_1}$ increases
from 0 to 0.2, and has a more shallow increase beyond that.  
It is not easy to confront this result with measurements from
the barred galaxies of the OSU and the Frei samples
\citep{Eskridge+00,FreiGGT96} for mainly two reasons. The first
  difficulty lies in the 
deprojection of the galaxies, since the value of the position and
inclination angles can be heavily influenced by the outer parts of the
arms. This is the well known Stock's effect \citep{Stock55} 
and may influence tightly wound spirals differently from open
ones. This could introduce a bias in the analysis and thus has to be
treated with particular care. The second difficulty comes from the
fact that the quantity $Q_{t,L_1}$
is not easy to obtain from observations since it implies
knowledge of the corotation radius and some information of the dark
matter content in this region.
 
Such a relation has not been found by density wave theory because this
has been fully developed only in the linear regime, so that a change in
the bar amplitude will only change the amplitude of the spiral (for
reviews, see \citealt{Toomre77} and \citealt{Atha84}).
It is, however, present in the response
simulations of \cite{Schwarz84}, as expected, since, as we showed in
Paper III the gas-response spirals and rings obey
manifold dynamics.

\section{Main ring properties}
\label{sec:ringprop}

\subsection{Frequency of ring types}
\label{subsec:frequency}

\citeauthor{Buta95} (\citeyear{Buta95}) classified 3692 galaxies in
his CSRG catalogue and found that about 53\% of them 
have an outer ring. 25\% of the galaxies with an outer ring have an
$R_1$ ring or $R_1^{\prime}$ pseudo-ring morphology. About 15\% have an
$R_2$ ring or $R_2^{\prime}$ pseudo-ring and 4\% an $R_1R_2$
ring morphology. The rest of the ringed galaxies have a single near
circular outer ring. 

Our model reveals what part of the parameter space corresponds to what
type of rings (see Paper III), but we can not translate this into
frequencies of observed ring types because our models are not exact
replicas of observed bars. It is possible, however, to
make a few relevant comments. In particular, in Fig. 4 of Paper III we
examined  the morphology of models with Ferrers bars, which are, to
date, the most realistic analytic bar model potential. For these, we
found that in most cases the manifold morphology is of
$R_1$, or $R_1^{\prime}$ type. This is in good agreement with the fact that
$R_1$ is the most frequently observed of the three ring types. 
Although this agreement is very welcome, it should not be
considered as a confirmation of our theory. 
On the other hand, it would have been worrying if
observations showed that $R_2$, or $R_1R_2$ rings were most common
since this would have implied a rather bizarre distribution of bars on the 
(bar strength, bar pattern speed) plane, and/or very strong differences
between Ferrers potentials and the potentials of observed bars.  

\subsection{Ratio of ring sizes}
\label{subsec:ratsize}

\subsubsection{General remarks}
\label{subsubsec:generalities}

As shown in Paper III, ring sizes depend strongly on the model
potential used. Moreover, our models are not exact representations of
observed bars and we neglect the self-gravity of the ring
structure. Therefore, quantitative comparisons of model and observed ring ratios
may well not show agreement. 

There is a further complication concerning the measurement of inner
ring sizes in models. As 
described in Paper III, weak bars have weakly unstable $L_1$ and $L_2$
Lagrangian points and, after a revolution around the galactic
centre, the inner manifold branches retrace the same path, thus forming the
inner ring. In these cases it is relatively easy to measure the length
of the major and minor ring diameters. On the other hand, strong bars
have strongly unstable 
$L_1$ and $L_2$ and the inner manifold branches cover a new path after
the first revolution around the galactic centre, retracing the same
loci only after a couple or a few revolutions (Papers III and V). One
of these loci could  
correspond to the ring, and the others to the outer parts of
the bar, or all could correspond to the bar. In
the latter case there will be no inner ring, while in the former one
it is not easy to find the ring outline. In principle, this should
be measured from the self-consistent superposition of the corresponding
manifold parts. This, however, is beyond the scope of this paper,
so in what follows we measured, whenever relevant, the size and shape
of the inner ring from the loci of the {\em first} revolution 
of the inner manifold branches. This is arbitrary and thus could lead
to considerable systematic errors. In particular, it could lead to a
considerable 
underestimate of the ring minor axis. This will be discussed further
below. 

Further reasons for discrepancies can stem from the
observations. Galaxies are observed as projected on the sky, i.e. are
not necessarily face-on, and thus the axial ratios can not be directly
compared to models. \cite{Buta95} gave statistical estimates of the deprojected
axial ratios from a careful analysis using a model and relying on
assumptions. The obvious and indisputable assumption is that the 
observed galaxies are oriented randomly with respect to the line of
sight. Nevertheless, since rings can not be seen in edge-on, or near
edge-on, discs, there is a bias against highly inclined cases, which 
is dealt with in the modelling by an approximation. Further
assumptions include the fact that both discs and bars are vertically
infinitesimally thin, that the ring shapes are well approximated by
ellipses. and that the intrinsic shape of the axial ratio distribution
is either top-hat with a narrow extent \citep{Buta86}, or Gaussian
\citep{Buta95}. Moreover, problems may also come from selection effects in
the catalogue, especially for rings with small diameters. Thus, the
errors stemming from observations can also be important. They will,
moreover, be more severe for inner than for
outer rings and will bring an artificial reducing of the number of
small axial ratio cases \citep{Buta95}.

Despite the above arguments, we attempt in this section not only
qualitative but also quantitative comparisons and estimates, but it
should be kept in mind that the latter may be affected
by non-negligible error, particularly in the case of inner rings. Let
us call $D_o$ and $d_o$ the lengths of the major and minor axes, 
respectively, of the outer ring and by $D_i$ and $d_i$ the same
quantities for the inner ring. The relevant quantities which we can use
to make comparisons with observations are $d_o/D_o$, $d_i/D_i$ and
$D_o/D_i$. 

\subsubsection{Further models}
\label{subsubsec:further-models}

\begin{figure*}
\centering
\hspace{0.5cm}
\includegraphics[scale=0.7,angle=-90.0]{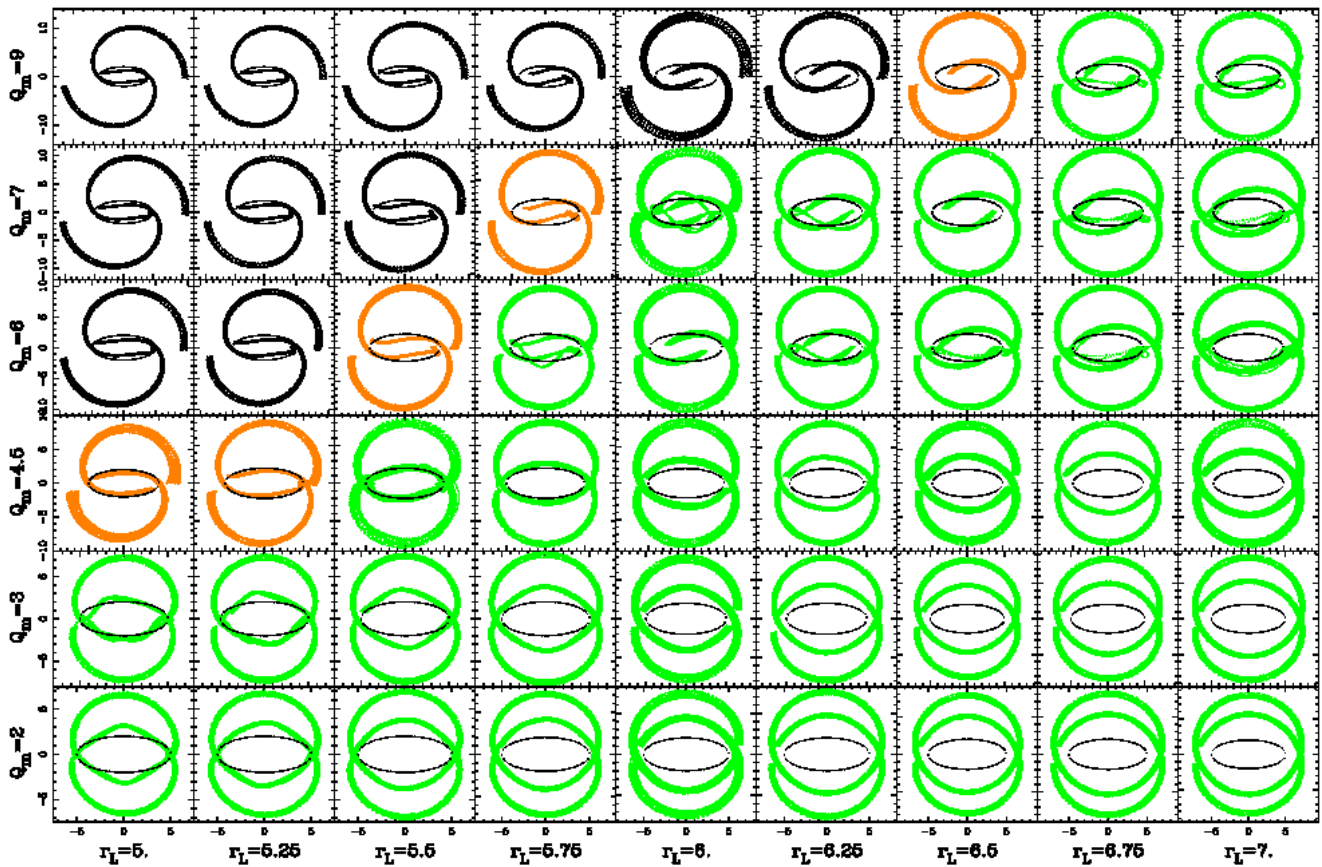}\hspace{0.5cm}
\caption{Effect of the bar strength and of the Lagrangian radius $r_L$
  on the manifold loci for a model with a Ferrers bar and a slightly
  decreasing rotation curve (see text). The strength of the bar is
  parametrised by the quadrupole moment $Q_m$.
  Each panel corresponds to a pair of ($Q_m$, $r_L$) values, given on
  the left (in units of 10$^4$) and on the bottom of the figure,
  respectively. The
  manifolds are plotted in a colour which is determined by their
  morphology: green for $R_1$, orange for $R_1'$ and black for
  spirals. The thin, black lines give the outline of the bar. 
  Compared to fig. 4 of Paper III, which has an
  identical bar model, this plot shows how a change in the
  axisymmetric component can affect the morphology of the response and
  the axis ratios of the rings. 
}
\label{fig:plot2Ddec}
\end{figure*}

\begin{table}
 \begin{center}
 \caption[]{Results for models with lenses: Axial ratio of the inner
   and outer ring sizes ($d_i/D_i$ and $d_o/D_o$, respectively) and
   $Q_{t,L_1}$ as a function of the relative  
   lens mass, $f_{lens}$ (see text for more information).}
 \label{tab:lens}
\begin{tabular}{ c c c c c c c}
  & \multicolumn{3}{c}{1} & \multicolumn{3}{c}{2} \\
 \cline{2-7} \\
$f_{lens}$ & $Q_{t,L_1}$ & $d_i/D_i$ & $d_o/D_o$ & $Q_{t,L_1}$ & $d_i/D_i$
& $d_o/D_o$  \\
 \hline
 \hline
 0.  & 0.077 & 0.51 & 0.76 & 0.077 & 0.51 & 0.76 \\
 0.2 & 0.079 & 0.57 & 0.80 & 0.087 & 0.53 & 0.75 \\
 0.5 & 0.062 & 0.67 & 0.81 & 0.076 & 0.55 & 0.75 \\
 1.  & 0.045 & 0.75 & 0.86 & 0.063 & 0.61 & 0.77 \\
\end{tabular}
\end{center}
\end{table}

\begin{table}
 \begin{center}
 \caption[]{Results for models with unstable or stable $L_1$ and $L_2$
   Lagrangian points: 
   Axial ratio of the inner and outer ring and 
   stability of the $L_1$ and $L_2$ as a function of the relative
   stabilising mass $F$ (see text and Paper III for more
   information on the models). }
 \label{tab:L12stable}
\begin{tabular}{ c c c c c }
$F$ & stability & $d_i/D_i$ &  $d_o/D_o$ & $D_o/D_i$\\
 \hline
 \hline
0.    &  unstable    &  0.81  &  0.90     &  1.50 \\
0.19  &  unstable    &  0.70  &  0.92     &  1.60 \\
0.20  &  unstable    &  0.72  &  0.91     &  1.71 \\
0.29  &  stable      &  0.64  &  0.85     &  1.91 \\
0.41  &  stable      &  0.60  &  0.85     &  2.06 \\
0.49  &  stable      &  0.55  &  0.84     &  2.07 \\
\end{tabular}
\end{center}
\end{table}

Since the ring shapes and the ratio of ring major axes diameters may
depend crucially on the adopted potential, we will, in the next few
subsections, consider not only models A, D and BW (see Paper III), but also
other potentials. 

To model galaxies with lenses, we 
add an extra Ferrers component which corotates with the bar and has
the same major axis as it \citep{Kormendy79}. The remaining parameters
were taken as for the fiducial model of Paper III, i.e. $a=5$,
$a/b=2.5$, $r_l=6$, $Q_m=4.5 \times 10^4$, $\rho_c=2.4 \times 10^4$ 
and $n=1$, all measured in the units given in Appendix~A of that
paper. The lens has an axial ratio $b/a$ = 0.9 and a mass
$M_{lens}=f_{lens}M_b$, where $M_b$ is the mass of the bar. The results
are given Table~\ref{tab:lens} for two such series of 
models. In the first one the total mass of the model was kept
constant (columns two to four, marked 1), so that a certain fraction of the bar
mass was converted into a lens. For the second one the lens
was simply added on to the model, without any change of the bar mass
(columns five to seven, marked 2). For the first series, increasing
the lens fraction 
decreases $Q_{t,L_1}$ (as expected), and renders the inner and outer
rings more axisymmetric. For the second series of models, adding a
lens makes a less clear-cut effect, because the total mass of the model
does not stay constant. It is, however, clear that it makes the inner
ring more axisymmetric and does not have a big effect on the shape of
the outer ring. 

We also considered models in which the $L_1$ and $L_2$ Lagrangian points
are stable. As in Paper III, this stability was achieved by adding
two mass concentrations in the form of Kuz'min/Toomre discs
\citep{Kuzmin56, Toomre63} each centred on one of the $L_1$ and $L_2$
Lagrangian points. In the 
examples given here we adopted for these discs a scale-length
$r_a=0.6$ and we tried several values of the fractional mass $F$,
where $F$ = $M_{KT}(r<r_a)/M_b(r<r_a)$ is the ratio of the mass of
one of the Kuz'min/Toomre disc ($M_{KT}$) to that of the bar ($M_b$), both
calculated within a radius equal to $r_a$ from their respective centres (see
Sect. 5 in Paper III for more information). These values, the
resulting stability and the measured outer ring axial ratios are
listed in Table~\ref{tab:L12stable}. We note that, as $F$ increases
past a certain threshold, the $L_1$ and $L_2$ turn from unstable to
stable. An increase in $F$ makes the inner ring considerably more
elongated, and has a smaller effect on the outer ring axial ratios,
which furthermore is not always monotonic. As a result, the ratio of
outer to inner major axes increases considerably with $F$. 

We also analysed the results for two further models, similar to model
A but with a different axisymmetric component.
For the first one we chose the axisymmetric potential leading to the 
circular velocity curve denoted by D in 
Paper II (see Fig. 7 in that paper.). This is slightly decreasing
in the regions around corotation and immediately beyond it. Thus,
$Q_{t,L_1}$ will be somewhat larger than for model A. We used the same
bar model and calculated the manifolds for the same grid of 
parameter values as for model A in Paper III. We call this model AD
and give the corresponding manifold morphologies in 
Fig.~\ref{fig:plot2Ddec}. Some notable differences are clear (compare 
Fig.~\ref{fig:plot2Ddec} of this paper with fig.~4 of Paper III), the
most important 
one being that there are many more spirals here than for the previously
used rotation curve. This is in agreement with the results and
discussions made so far, since higher $Q_{t,L_1}$ values lead to spiral
morphologies and lower ones to an $rR_1$ morphology (Paper III and
Sect~\ref{subsec:strengthmorpho} of the present paper). 

We also tried another axisymmetric component, such that the peak of the
circular velocity curve is nearer to the galactic centre, i.e. more adequate
for early type galaxies, and we denote the corresponding model by
AE. This introduced relatively little change to the ring
shapes, as will be discussed in the following sections.

\subsubsection{Outer ring axial ratios}
\label{subsubsec:outer-ratios}

\begin{figure}
\centering
\includegraphics[scale=0.33,angle=-90.0]{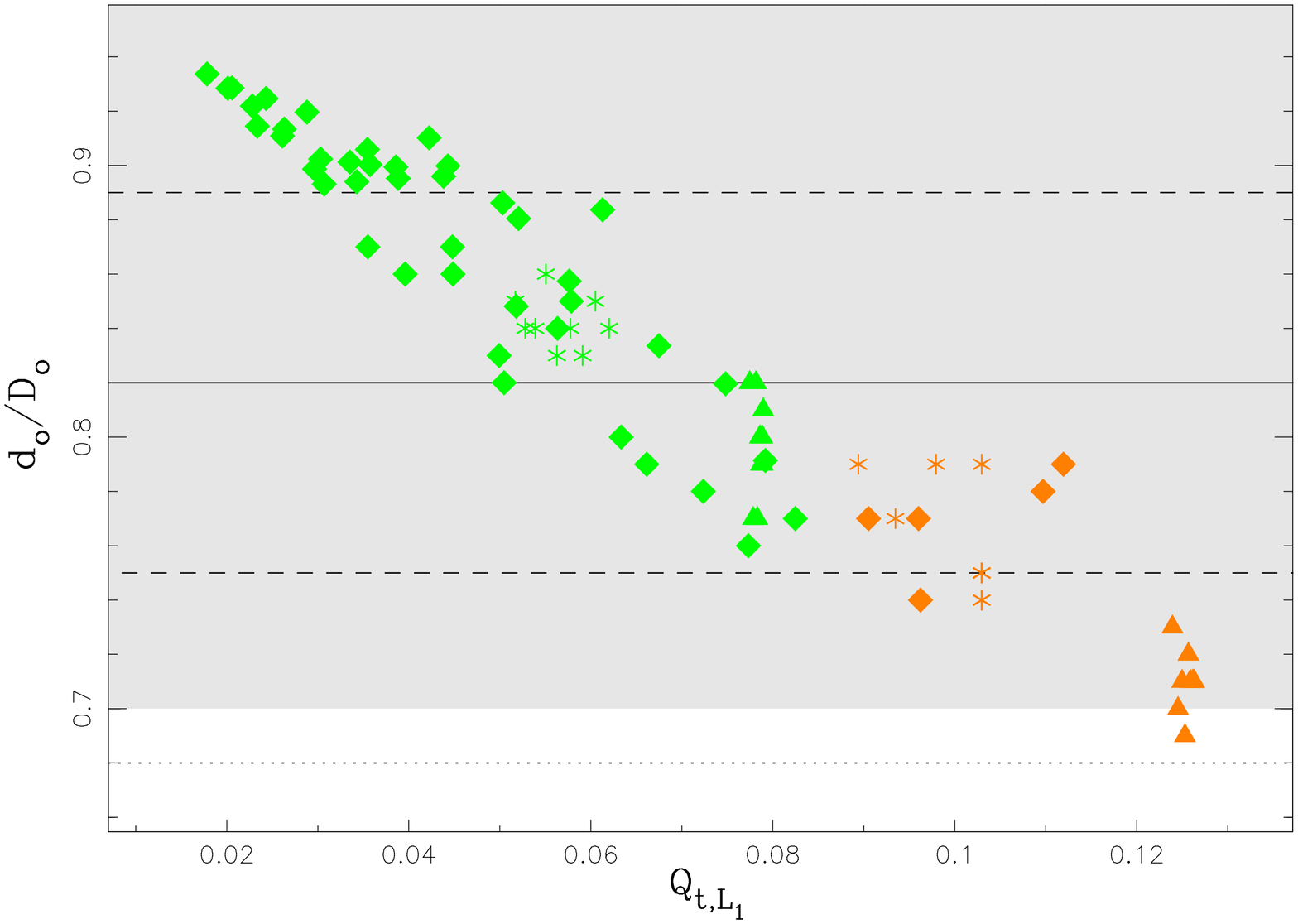} 
\includegraphics[scale=0.33,angle=-90.0]{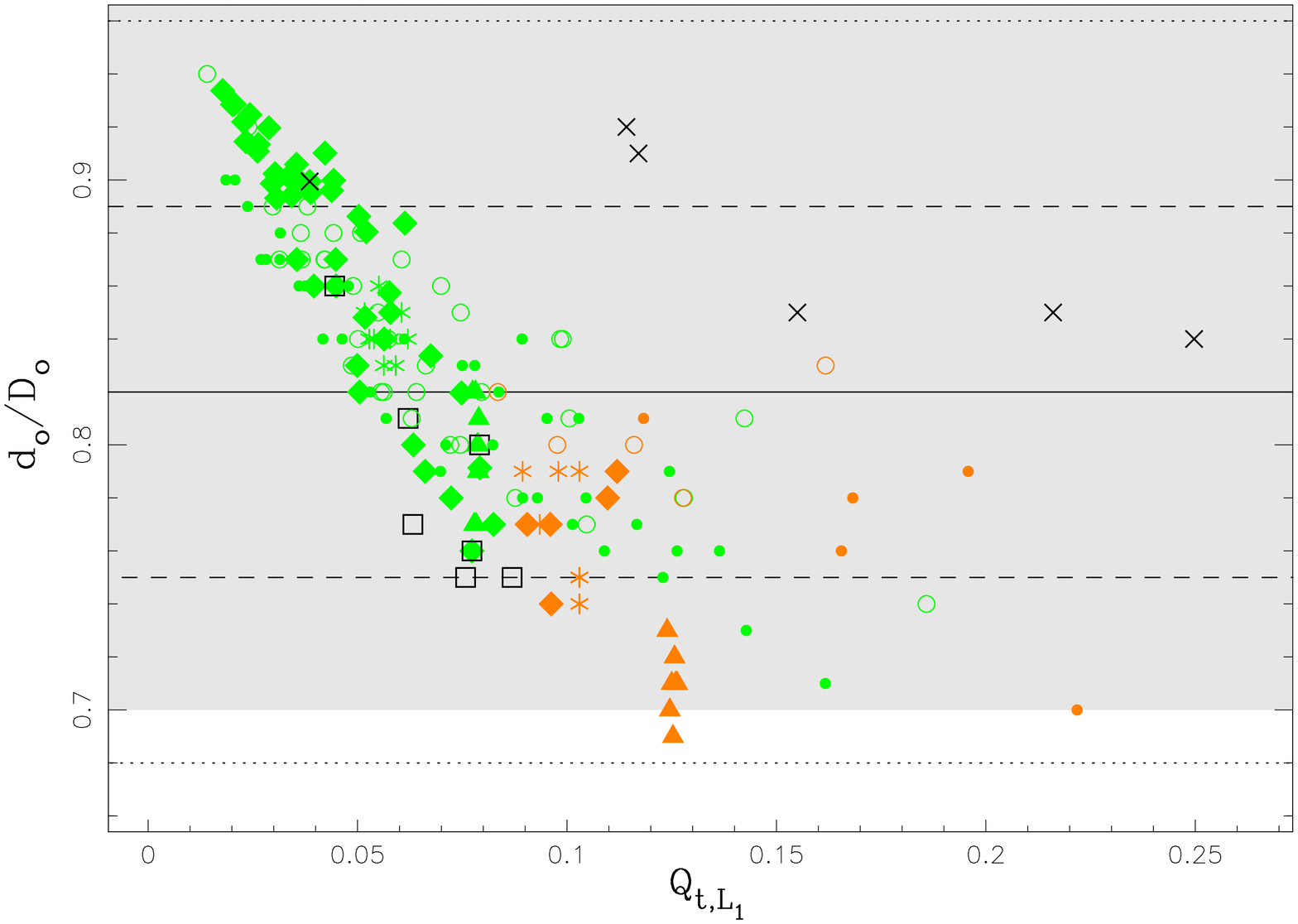} 
\caption{Relation between the strength of the bar in the corotation
  region, measured by $Q_{t,L_1}$, 
  and the axial ratio of the outer $R_1$, or $R_1'$ ring for our models.
  As in Paper III, different colours stand for different morphologies:
  green for $R_1$ and orange for $R_1'$.   The upper
  panel includes only morphologies for models A, D and BW 
  and shows a strong correlation, in the sense that galaxies with
  stronger bars have also more elongated outer rings. 
  Results from models A are given by filled diamonds,
  from models D by asterisks and from models BW by filled
  triangles. In the lower panel we add more models. 
  Models AD are plotted with a
  filled circle and models AE with an open one. Models with lenses are plotted
  with an open square  
  and models from Table~\ref{tab:L12stable} with an x sign. 
  The horizontal lines give statistical information on the 
  observed outer ring axial ratios, namely the mean (solid line) as
  well as the one $\sigma$ 
  (dashed) and the two $\sigma$ (dotted) deviations from it, as
  calculated for barred 
  galaxies from the CSRG \citep{Buta95}. The grey area shows the region
  of values found from the SCG survey \citep{Buta86}. Note that the 
  outer ring axial ratio values we find from our models are in the right range.
}
\label{fig:d0overD0-barstrength}
\end{figure}

Analysing the results from his CSRG catalogue,
\cite{Buta95} finds that the outer rings have a significantly
oval intrinsic shape, with an average intrinsic axial ratio of $<q_o>
= 0.87 \pm 0.14$. For his SB sub-sample alone, he finds
$<q_o> = 0.82 \pm 0.07$. He also finds that the
outer rings are rounder in SA than in SB galaxies, with an average
intrinsic axial ratio of 0.97 $\pm$ 0.21.  Applying the same procedure
to the Southern Galaxy Catalogue (SGC) he finds $<q_o> = 0.83 \pm
0.01$ \citep{Buta95}, while in a 
previous study with a somewhat different procedure he finds
that the preferred axial ratio range is 0.7 -- 1
for SB galaxies and 0.8 -- 1 for SAs and SBAs \citep{Buta86}.
There is thus general agreement between the various estimates of the
observed outer ring shapes.

Amongst the roughly 300 models discussed in Sect.~3 of Paper III and
in Sect.~\ref{subsubsec:further-models} above, 180 have an $R_1$ or
$R_1'$ morphology. This 
number is sufficiently large for us to be able to make some
analysis. We therefore measured the major and minor axes of the outer ring
for this sub-sample and plot their ratio as a function of
$Q_{t,L_1}$ in Fig.~\ref{fig:d0overD0-barstrength}. 
We have too few $R_2$ models for any reasonable analysis and
this morphology is also more rare in real barred galaxies. Finally   
we can not compare the properties of $R_1R_2$ rings with observation,
because the measurements are 
not uniquely defined since it is not clear whether one should measure
the $R_1$, or the $R_2$ part. \cite{Buta95} measures both and then takes an
average of the values weighted according to the strength of each
feature. We can not do this for our models because the manifold
calculations we presented so far can give us information 
on the existence and shape of a feature, but not on how much
material is trapped by it, the latter depending on the evolutionary
history of the galaxy. We thus have no information on how
strong the $R_1$ and $R_2$ are. In order to obtain this information,
we would need to follow the trapping by the manifold during the disc
formation and evolution stage. A simplified version of such a
calculation has been given in Sect.~\ref{subsec:trailing}, but a more
realistic version would necessitate considerable knowledge on how the
disc and its main substructures were formed. 
 
The horizontal lines in Fig.~\ref{fig:d0overD0-barstrength} give the
average value and the one and two sigma deviations from it for barred
galaxies from the CSRG \citep{Buta95}. The grey area shows the region
of values found from the SCG survey \citep{Buta86}. We note that the 
values we find from our models are in the right ballpark. In fact, all
values are within two sigma from the observed mean and most models are
within one sigma from it. Adding results from other types of models (with
different axisymmetric components, or lenses, or stable $L_1$ and $L_2$
Lagrangian points -- see Sect.~\ref{subsubsec:further-models}) does not
change in any qualitative way these results (lower panel of
Fig.~\ref{fig:d0overD0-barstrength}). Concerning the outer ring shape,
there is thus 
very good agreement between theory and observations, which holds for
all the very wide range of potentials we used.  

The upper panel of Fig.~\ref{fig:d0overD0-barstrength} shows a tight
correlation, in the sense that galaxies with stronger bars have also
more elongated outer rings. Most of the points on this plot correspond
to A models, with some D and BW models. In fact, it would have been
rather difficult to assert the existence of this correlation from the
D or from the BW models separately, because their $Q_{t,L_1}$ values
are rather clustered. Nevertheless, both the D ad BW models fall well on the
correlation outlined by the A models, thus reinforcing it.

Let us now check whether this correlation still holds for a yet wider
set of model potentials (lower panel of
Fig.~\ref{fig:d0overD0-barstrength}), namely all those introduced in  
Sect.~\ref{subsubsec:further-models}. All models with lenses fall neatly
on the correlation. Most of models with a AD or AE axisymmetric
component fall also on or very near the correlation. Only four points
(out of 180) are considerably off. Furthermore, three of these correspond to
marginal cases, i.e. cases which could be classified either as $R_1'$, or
as spiral. Thus we can conclude that, although the correlation is widened
and could be called a trend, it is certainly still present, even with
the very wide variety of the models used. This figure also gives an
indication that, for this trend at least, the axisymmetric part of the
potential, i.e. the circular velocity curve, makes more difference
than the analytical expression used for the bar potential. Considerable
more work is, however, necessary in order to assess this.

On the other hand, models with stable $L_1$ and $L_2$ Lagrangian
points definitely stand apart from the rest and are off the
correlation although they have realistic values of the outer ring
axial ratio. 
It is unclear at this point whether this
argues only that this correlation will not be followed by models with
stable $L_1$ and $L_2$ Lagrangian points, or, more strongly, that the
correlation will be followed only by certain types of model potentials. 

\begin{figure}
\centering
\includegraphics[scale=0.33,angle=-90.0]{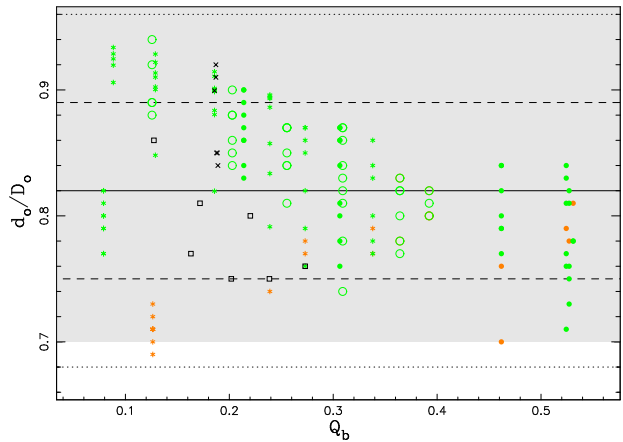} 
\caption{Outer ring axial ratio as a function of the bar strength. The
  latter is measured by $Q_b$ (see text for definition). Models A, D
  and BW are plotted by asterisks. The remaining symbols, and the
  colours, lines and shading are as in the lower panel of
  Fig.~\ref{fig:d0overD0-barstrength}. 
}
\label{fig:d0overD0-Qb}
\end{figure}

It would be interesting to check whether the correlation we find in
Fig.~\ref{fig:d0overD0-barstrength} is also valid for 
{\em observed} $R_1$ and $R_1'$ outer rings. It is in agreement with the
average values found by \cite{Buta95}, since these show that barred
galaxies have more elongated outer rings than non-barred
ones. Unfortunately, it is not possible to make any more accurate tests
with only the CSRG data, because it is necessary to deproject each galaxy
individually before measuring the ring axial ratios. Furthermore, it is
necessary to use an appropriate measure of the bar strength. The
correlation in Fig.~\ref{fig:d0overD0-barstrength} uses $Q_{t,L_1}$
which measures the non-axisymmetric forcing at corotation. Any such
measure at this radius, or somewhat beyond should be
appropriate. Measures at smaller radii, however, may not be. To check
this further, we replotted the data of the lower panel of
Fig.~\ref{fig:d0overD0-barstrength}, now using $Q_b$ as a bar strength
measure. This quantity is very often used in both theoretical and observational
studies (e.g. \citealt{ButaBK03, ButaLS04, LaurikainenSB04,
  LaurikainenSBV04, DurbalaBSVM09} and references therein). It relates
to the maximum non-axisymmetric forcing and is defined as the maximum
of $Q_t(r)$ in the bar region. 

The result is shown in Fig.~\ref{fig:d0overD0-Qb}. We have
not included the results from the Dehnen models, because this ad hoc
potential makes it difficult to measure $Q_b$ for the parameters we
are using. Even without these points, however, there is no clear
correlation. The best correlation is obtained if we look at 
models A alone and exclude all others.
For this case, the rank correlation coefficient is -0.81,
while the same models when plotted as a function of $Q_{t,L_1}$ give a
rank correlation coefficient of -0.94. Thus the correlation is less
tight, but the trend itself is not lost. On
the other hand, if we include also the BW models, then the rank
correlation coefficient becomes -0.001 for $Q_b$ and -0.95 for
$Q_{t,L_1}$, so that the correlation is quite strong for
$Q_{t,L_1}$ and is totally lost for $Q_b$. Thus, whether it is possible to
use $Q_b$ instead of $Q_{t,L_1}$ depends on the comparison at hand. In
general, for properties related to the bar or to the inner manifolds
$Q_b$ should be a good quantity to use. For the properties of the spirals and
outer rings, however, the use of $Q_b$ may be hazardeous. In
the absence of any further information it is safer and preferable to avoid
the use of $Q_b$ for comparison with observations related to the
spirals and rings. Nevertheless, $Q_{t,L_1}$ is a more difficult quantity to
obtain from observations than $Q_b$, since it presumes an albeit rough
knowledge of 
the Lagrangian radius $r_{L}$. In Paper V we discuss how this
quantity can be obtained with the help of our theory and how reliable
this estimate can be in various cases. One can thus conclude that such
comparisons with observations should be done with great care since
the value of $Q_{t,L_1}$ may contain some error due to a wrong
estimate of $r_{L}$, while $Q_b$ may give wrong estimate of
$Q_{t,L_1}$ and therefore of the quantities that depend on it.

\subsubsection{Inner ring axial ratios}
\label{subsubsec:inner-ratios}

\begin{figure}
\centering
\includegraphics[scale=0.33,angle=-90.0]{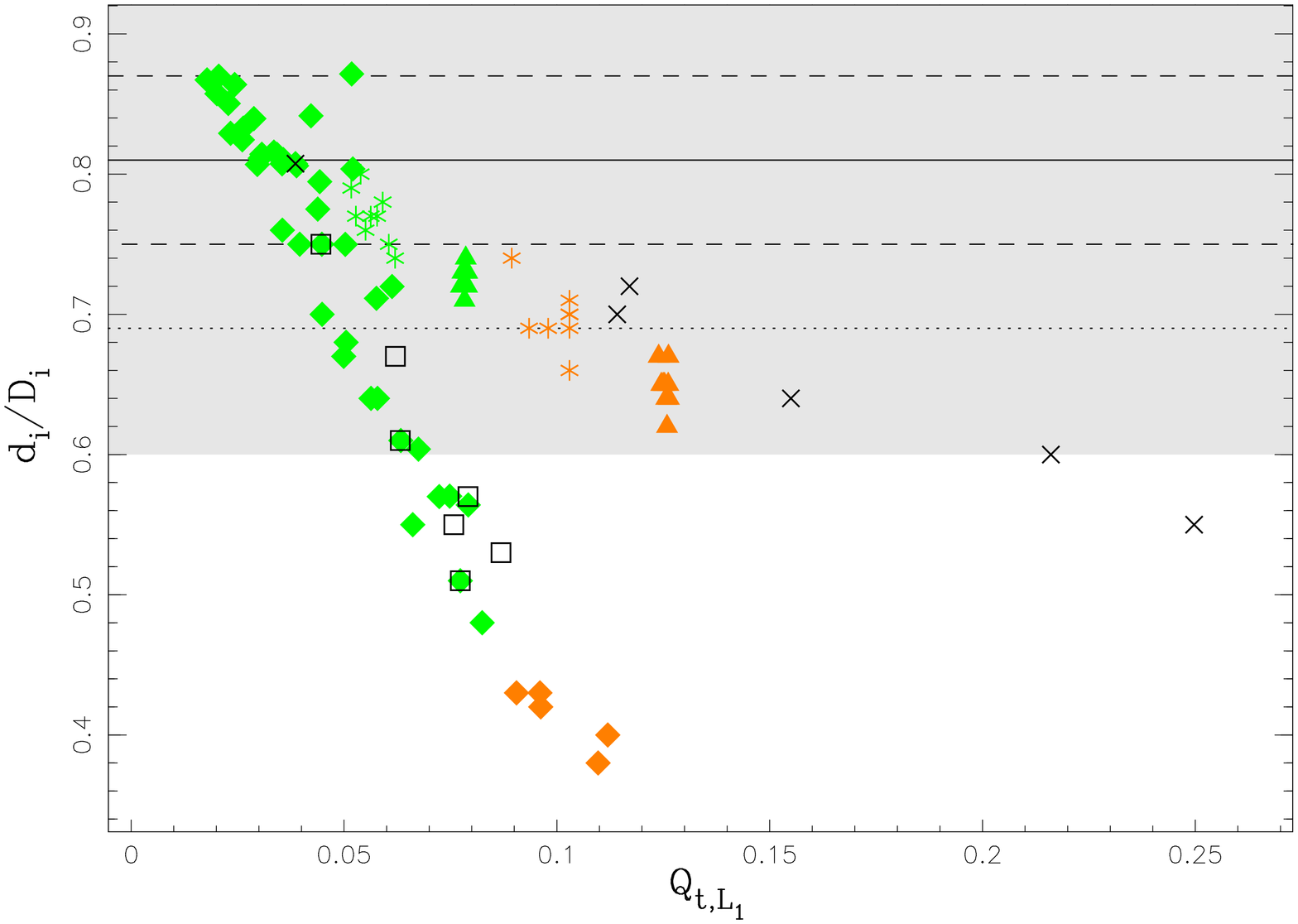} 
\includegraphics[scale=0.33,angle=-90.0]{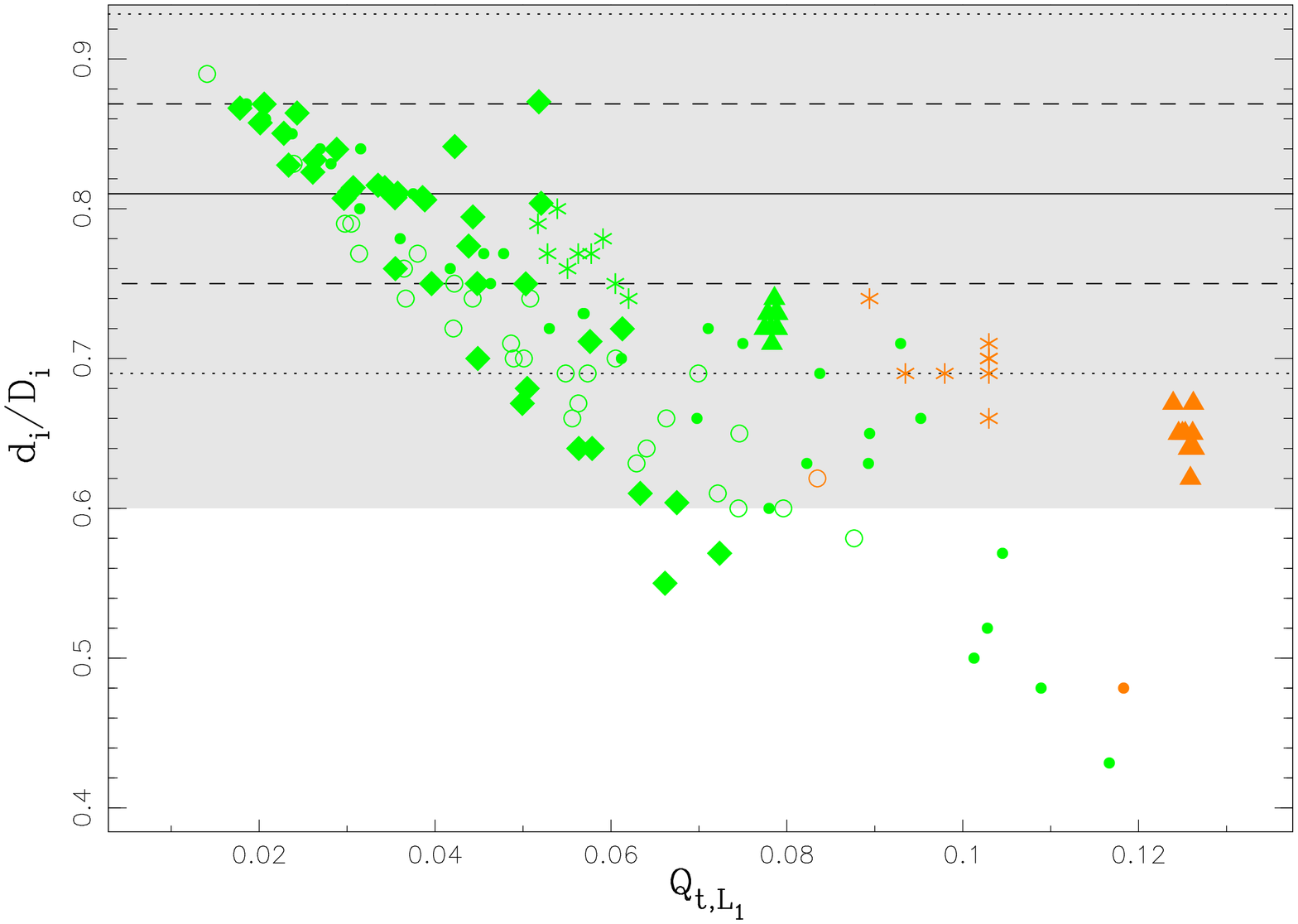} 
\caption{Relation between the strength of the bar near corotation, measured by
  $Q_{t,L_1}$, and the axial ratio of the inner ring for our models.
  There is a strong correlation, in the sense that galaxies with
  stronger bars have also more elongated inner rings. The horizontal
  lines and the gray areas give some statistical information on the
  corresponding 
  observed quantities (see text). Colours and symbols are as in
  Fig.~\ref{fig:d0overD0-barstrength}, upper and lower panels,
  respectively.  
}
\label{fig:dioverDi-barstrength}
\end{figure}

\cite{Buta95} analysed the inner ring data from the CSRG in a similar
way as for the outer rings. He finds that the inner rings have a
significantly oval intrinsic shape and in fact are on average more
elongated than outer rings. For his SB sub-sample he finds $<q_o> = 0.81 \pm
0.06$, where $<q_o>$ is the average intrinsic axial ratio. As discussed
above for outer rings, he can not seek correlations of the ring axial
ratio with galaxy, or bar parameters because this would necessarily
imply deprojecting each galaxy individually. He does, however, find
that inner rings are rounder in SA 
than in SB galaxies with an average intrinsic axial ratio of 0.92
$\pm$ 0.09 for the former. Applying the same procedure to the SGC,
\cite{Buta95} finds $<q_o> = 0.82 \pm 0.02$, while in a 
previous study with a somewhat different procedure \cite{Buta86} finds
that the axial ratios are uniformly spread in the range 0.60 -- 0.95
for SB galaxies and 0.85 -- 1 for SAs. 

Fig.~\ref{fig:dioverDi-barstrength} gives the same information as
Fig.~\ref{fig:d0overD0-barstrength}, but now for inner rings. Results
from model A (upper panel) show a very tight correlation between the inner ring
axial ratio and the $Q_{t,L_1}$ measure of the bar strength. Models
with lenses also lie on the same correlation. Models D and BW (upper
panel) also show a correlation, which is followed by the models from
Table~\ref{tab:L12stable}. The regression lines of these two 
correlations, however, do not coincide, the former being much steeper
than the latter. 

A further point which is clear from the upper panel of
Fig.~\ref{fig:dioverDi-barstrength} is that some of the models have an
inner ring shape which is much more elongated than the observations.  
Several arguments need to be taken into account here. It should be kept
in mind that all the biases, both observational and theoretical, tend,
as discussed in Sect.~\ref{subsubsec:generalities}, to increase
this discrepancy and their cumulated effect could be considerable. 
Amongst the most important ones is the fact that, as we already
mentioned, the inner manifold branches can correspond either to the
ring or to the outer parts of the bar. For models A, AD and AE where the bar
outline is known, it is possible to distinguish between the
two. Unfortunately this is impossible for models D and BW and not
straightforward for models from Table~\ref{tab:lens} or 
Table~\ref{tab:L12stable}. We examined all our A, AD and AE models and
eliminated those for which the inner manifolds are more likely to
correspond to the bar than to an inner ring. This exercise is not
trivial, because our calculations are not self-consistent, i.e. the
gravity from the manifold itself is not taken into account. If the
manifold locus is somewhat outside, but near the bar outline, then in real
galaxies, or in self-consistent calculations, the bar outline could 
somewhat change due to the gravitational pull from the manifold and,
provided the distance between the two is small, the two could merge. 
This is something we can not take into account with our calculations. 
We thus eliminated only the cases where we estimated from our
non-self-consistent 
simulations that the manifold pertained to the bar more than to the
inner ring. We then plot the remaining models, including all possible
families, in the lower panel of Fig.~\ref{fig:dioverDi-barstrength}.
 
Note that most of the models are in the area outlined by the
observations, but a handful are still clearly outside. We examined by
eye these cases and found that they are borderline, i.e. perhaps a
self-consistent calculation would lead us to classify them as contributing
to the outer part of the bar, rather than to the ring. 

We also examined which potentials increase this discrepancy, i.e. make
more elongated inner rings, and which alleviate it. First considering
the rotation curves, we see that AD rotation curves create more
elongated rings, i.e. increase the discrepancy between models and
observations. This can be understood since in AD models the
axisymmetric force is relatively smaller over some of the relevant
radial extent than in A models and this, according to our previous
results, should lead to more elongated rings, as is indeed found. For
models AE the difference is less clear-cut. We can, however, see that
for the largest $Q_{t,L_1}$ values, the AE have less elongated rings
than the A models, thus alleviating the discrepancy. This can be
understood by the fact that, for the most elongated rings, the inner
manifold branches reach radial regions where the axisymmetric force is
relatively smaller in the AE models than in the A ones. This would
then lead to less elongated inner rings in the cases with largest
$Q_{t,L_1}$, as indeed found in our calculations.

The problem of too elongated inner rings was already
encountered by \cite{Schwarz85}, who attributed it to the potential he
was using. To solve it he proposed to include a lens component.  
Fig.~\ref{fig:dioverDi-barstrength} and Table~\ref{tab:lens} show that
this is indeed a possible solution. For example, the inner ring axial
ratio of one of the A models, which without a 
lens is equal to 0.51 (i.e. nearly three sigma
below the average), becomes 0.75 (i.e. within one sigma
from the mean) if half of the bar mass is turned into a lens. This of
course can not be the only solution, since not all galaxies with inner
rings have lenses.

The above argue convincingly that the axial ratio of the inner ring
depends on the potential used, stronger relative non-axisymmetrc
forcings leading to more elongated rings. The results we get with our
potentials are in relatively good agreement with observations and an
excellent agreement could be reached if the non-axisymmetric forcings
in the models were somewhat lower in the region covered by the inner
ring.  

\begin{figure}
\centering
\includegraphics[scale=0.33,angle=-90.0]{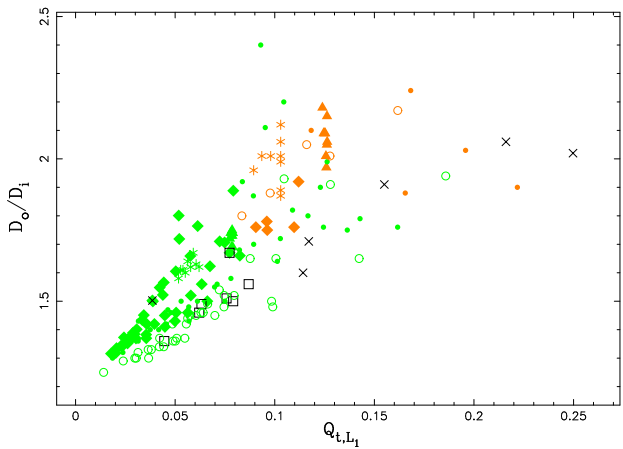}
\caption{Ratio of the major axis of the outer ring to the major axis
  of the inner ring as a function of $Q_{t,L_1}$ for all our models.
  Green symbols denote measurements from $R_1$ models, and orange ones
  from $R_1'$ models. The symbol used depends on the model potential.
  Results from models A are given by diamonds,
  from models D by asterisks and from models BW by filled
  triangles. Models AD are plotted with a filled circle and models AE
  with an open one. Models with lenses are plotted with an open square  
  and models from Table~\ref{tab:L12stable} with an x sign. 
}
\label{fig:DoDiQt}
\end{figure}

\subsubsection{Statistics on the major axis ratios}
\label{subsubsec:major-ratios}

\begin{figure}
\centering
\includegraphics[scale=0.33,angle=-90.0]{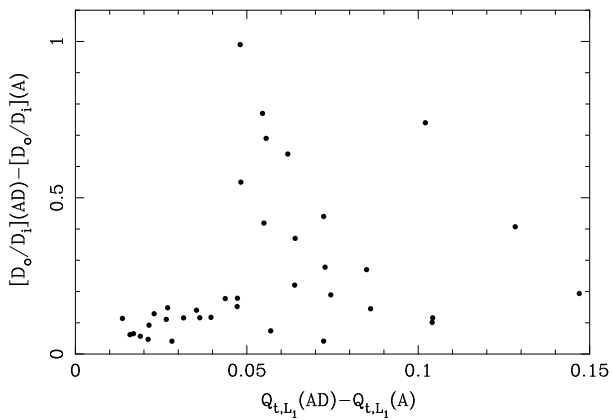}\hspace{0.5cm}
\caption{The effect of a change of the model circular velocity curves
  on the ratio of the outer to inner major axes. The bar parameters do 
  not change, so that each point shows the difference between two
  models with the same bar, but different circular velocity curves. On
  the abscissa we plot the difference between $Q_{t,L_1}$ for model AD
  and the same quantity for model A. On the ordinate we plot the
  difference between the ratio of the outer to inner major axes,
  $D_o/D_i$, for models AD and the same quantity for model A. We note that the
  change from A to AD increases both the value of $Q_{t,L_1}$ and of $D_o/D_i$,
  thus bringing the latter measurements nearer to the observed values.
}
\label{fig:DoDiQt-arrows}
\end{figure}

Analysis of the ratio of outer to inner ring sizes in observed
galaxies has shown that the corresponding histogram is rather concentrated to
values around 2 \citep{AthaBCS82, Buta86, Buta95}. This is a
clear-cut result and it would be nice to test our models with
it. However, as already discussed in detail in 
Sect.~\ref{subsubsec:generalities}, this is a very difficult
comparison to make, since the ratio in question is strongly
affected by errors and biases due to both difficulties in measuring
the observed and the model ring sizes and to the fact that
measurements in observations have not been done in individually
deprojected galaxies. Finally, it should be
remembered that we can compare the range of values covered by
observations, but we can not compare the corresponding 
distributions, since the models we have presumably do not cover the parameter
space in the same way as the observed galaxies. In fact the observed 
distribution of some of the relevant parameters is still largely
unknown, although considerable progress has been made for others
\citep{Gadotti09}. We will, nevertheless, attempt a general discussion
and comparison with observations and try and outline which types of
potentials give the best (or the worst) results. 

Fig.~\ref{fig:DoDiQt} shows the ratio of the outer to inner ring major
axes, $D_o/D_i$, plotted as a function of $Q_{t,L_1}$. It is clear that
although a large number of our models have values in the same ballpark
as the observations, many have considerably smaller
values. Fig.~\ref{fig:DoDiQt} also shows a trend which, although it is
not sufficiently tight to be called a correlation, is still very
clear. Namely, the ratio $D_o/D_i$ is smaller in cases with smaller
$Q_{t,L_1}$ values.  Fig.~\ref{fig:DoDiQt} also gives valuable
information on which 
types of potentials are in better agreement with the observations. 

{\em All} models with an $R_1'$ morphology have a ring ratio in agreement
with observations, independent of the type of potential used. The same
is true for nearly all models of the AD type, as well as for
models whose $L_1$ and $L_2$ Lagrangian points are stable.
All these give results which
are roughly in the right range of values. On the other hand, many of the
remaining cases give smaller values of the axial ratio,
i.e. outside the desired region. Such results come from models with an
$rR_1$ morphology with potentials A, or AE.

To make the effect of the circular velocity curve clearer we compare
in Fig.~\ref{fig:DoDiQt-arrows} models A to models AD. For this we
consider pairs of models having the same bar parameters, but one
having an A circular velocity curve and the other an AE one, and
calculate the difference in $Q_{t,L_1}$ and in the
ratio of outer to inner ring major axes, $D_o/D_i$. We see that both
quantities are larger for AD models and that the difference can be
considerable. Thus such a change would bring our $D_o/D_i$ in better
agreement with observations.   

This clearly underlines the effect of the properties of the model potentials on
quantitative measures and fully justifies the corresponding arguments
advanced in Sect.~\ref{subsubsec:generalities}. It should {\em not},
however, be interpreted as arguing that
rotation curves should be slightly decreasing at and
beyond corotation. It argues instead for somewhat larger $Q_{t,L_1}$     
values in that region, something that could be also due to a somewhat larger
non-axisymmetric forcing. This would be the case in self-consistent
treatments, since the potential from the non-axisymmetric rings would
be included, and also if the bar model potential dropped less abruptly
with radius that the $n$ = 1 Ferrers models we use. Thus, such an increase
seems very natural and it would easily lead to values of the ring size
ratios in good agreement with observations. 

\section{Summary}
\label{sec:summary}

In previous papers (Papers I, II and III) we proposed a theory to
explain the formation and properties of rings and spirals in barred
galaxies. In this paper we compared their morphological properties to
that of observed ones. Our theory can produce both spirals and inner
and outer rings. Furthermore we get all three observed types of outer
rings , i.e. $R_1$, $R_2$ and $R_1R_2$ rings and 
pseudo-rings, and no other morphologies. This means that all outer rings
produced  by our theory have a qualitatively realistic morphology and
that all observed morphologies are qualitatively reproduced. We also
reproduce the dimples often seen clearly in $R_1$ rings. Furthermore, 
we predict that if the non-axisymmetric forcing is relatively low the
resulting morphology will be an $R_1$ ring, while if it is stronger, it 
will be a spiral. 

Spirals due to this theory start off from a location not far from the
end of the bar and are trailing, in good agreement with
observations. They are in the vast majority of cases two armed,
although it is possible, if specific potentials are used, to have arms
of higher multiplicities. Their shape is also realistic and we made
a successful quantitative comparison with the arms of NGC 1365. They
are able to reproduce the ``fall back'' of the arm towards the more
inner region,
seen in NGC 1365 and in many other barred spirals, something which
density wave theory could not achieve. We also predict that stronger
non-axisymmetric forces in the region around and somewhat beyond
corotation will result in more open spirals. 

Quantitative comparisons are less straightforward to make, not because
they imply tighter constraints, which they do not, but because they
involve not only the theory, but also the potential to which it is
applied. We are confident that our theory is dynamically correct, and
the comparisons with observations made here and in Paper V make us
optimistic about its applicability to galaxies with spirals or
rings. On the other hand, we are aware of the shortcomings of our
potentials. First and foremost, our analysis is not self-consistent, so that we
neglect the self-potential of any structure they create. Furthermore, the
Ferrers' potentials, although the best of all the available analytical
bar potentials, have a number of shortcomings, the most important one
for our particular application being that the force they create diminishes
too rapidly at large radii. The other potentials we used,
although they lack this particular drawback, are ad hoc, and
therefore not necessarily realistic.
The above remarks should be kept in mind when assessing our
quantitative comparisons. Our aim in such cases was to find
whether the potentials we used give results in agreement with
observations, or, in case of disagreement, whether any obvious and
straightforward changes of the potential can bring agreement. 

For the outer ring our
analysis concentrates on $R_1$ cases, for which we have sufficient
models to reach safe conclusions. With all the potentials used so far, the outer ring shape is in good
agreement with observations. Most models are within one $\sigma$ from
the observed mean and all are within two $\sigma$. We tried different
changes of the potential -- such as different rotation curves, adding
a lens, or stabilising the $L_1$ and $L_2$ Lagrangian points (see
Paper III) -- and this result remained unchanged. There is thus
excellent agreement between model and observed outer ring shapes.

For the inner ring,
we found several models with inner ring shapes more elongated than
observed. Examining by eye these cases, we found that for most, if not
all, the inner manifolds correspond more to the outer regions of the
bar and not to the inner ring. We can thus conclude that there is a
fair agreement between theory and observations regarding the inner
ring axial ratio. 

Finally, the ratio of outer to inner ring major axes
gives good agreement for some types of models, but not for all. Thus
all pseudo-ring cases ($rR_1'$), independent of the model potential, give 
good agreement. This is the case also for both $rR_1$ and $rR_1'$
morphologies in models with a circular velocity curve showing a slight
decrease at 
or somewhat beyond corotation, for models with stable $L_1$ and $L_2$
Lagrangian points and for a number of other models.
On the other hand, agreement for cases with an $rR_1$ morphology and A
or AE rotation curves (with or without a lens) often have too small
a value of this ratio.  
It is thus clear that the underlying potential plays an important
role. Our calculations suggest that models such that
$Q_{t,L_1}$ is somewhat decreased within corotation and increased
outside it would give good agreement for all ring size measurements,
namely the inner and outer ring shapes (ratio of minor to major axes)
and the ratio of the outer to inner ring major axes.  
It is, however, beyond the scope of this
paper to find a model achieving this at all relevant radii, thus
includes the best parts of all the models considered, 
particularly since this exercise may prove futile if
self-consistency is taken into account.

We find
a trend between the ring axial ratio and the relative amplitude
of the non-axisymmetric forcing in the region around, or immediately
beyond corotation, in the sense that stronger bars are linked with
more elongated rings. This is true for both inner and outer
rings. There is also a trend between the ratio of
outer to inner ring major axes and this measure of non-axisymmetric
forcing. If we focus on only one model family, or a couple of them, these
trends can become clear correlations, but including all models
broadens them, sometimes substantially. 

In Paper V, the next and last paper of this series, we will present a
number of other comparisons with observations and predictions,
including kinematics, bar shape, radial drift and abundance gradients.
We will discuss pattern speed prediction and give a global
appreciation of the application of our theory to observed spirals and
rings. We will also consider the effects of time evolution and of
self-consistency. Finally we will discuss other theories and
simulations and how they link to this work.

\section*{Acknowledgements}

EA thanks Scott Tremaine for a stimulating discussion on the manifold
properties. We also thank Ron Buta for useful discussions and email 
exchanges on the properties of observed rings and for allowing us to
reproduce images of two galaxies in Fig. 2 and the
anonymous referee for a careful reading of the manuscript. This work was partly
supported by grant ANR-06-BLAN-0172 and by the Spanish MCyT-FEDER Grant 
MTM2006-00478. This paper made use of two images from the Digital Sky
Surveys which were produced at the Space Telescope Science Institute
under the U.S. Government Grant NAG W-2166. 
\bibliography{manifIIa_after_ref}

\label{lastpage}

\end{document}